\documentclass[twocolumn,showpacs,preprintnumbers,amsmath,amssymb]{revtex4}
\usepackage{graphicx}
\usepackage{dcolumn}
\usepackage{bm}
\usepackage{color}

\usepackage{epstopdf}
\begin{document}
\newcommand{\kvec}{\mbox{{\scriptsize {\bf k}}}}
\newcommand{\lvec}{\mbox{{\scriptsize {\bf l}}}}
\newcommand{\qvec}{\mbox{{\scriptsize {\bf q}}}}
\newcommand{\pvec}{\mbox{{\scriptsize {\bf p}}}}
\def\eq#1{(\ref{#1})}
\def\fig#1{\hspace{1mm}\ref{#1}}
\def\tab#1{\hspace{1mm}\ref{#1}}
\title{Pseudogap in Eliashberg approach based on electron-phonon\\ and electron-electron-phonon interaction}
\author{R. Szcz{\c{e}}{\`s}niak$^{\left(1,2\right)}$}
\email{szczesni@wip.pcz.pl}
\author{A. P. Durajski$^{\left(1\right)}$}
\email{adurajski@wip.pcz.pl}
\author{A. M. Duda$^{\left(1\right)}$}
\email{annduda87@gmail.com}
\affiliation{$^1$ Institute of Physics, Cz{\c{e}}stochowa University of Technology, Ave. Armii Krajowej 19, 42-200 Cz{\c{e}}stochowa, Poland}
\affiliation{$^2$ Institute of Physics, Jan D{\l}ugosz University in Cz{\c{e}}stochowa, Ave. Armii Krajowej 13/15, 42-200 Cz{\c{e}}stochowa, Poland}
\date{\today}
\begin{abstract}
The properties of the superconducting and the anomalous normal state have been described by using the Eliashberg method. The pairing mechanism has been reproduced with help of the Hamiltonian, which models the electron-phonon and electron-electron-phonon interaction (EEPh). The set of the Eliashberg equations, which determines the order parameter function ($\varphi$), the wave function renormalization factor ($Z$), and the energy shift function ($\chi$) has been derived. It has been proven that for the sufficiently large values of EEPh potential, the doping dependence of order parameter ($\varphi/Z$) has the analogous course to that observed experimentally in cuprates. The energy gap in the electron density of states is induced by $Z$ and $\chi$ - the contribution from $\varphi$ is negligible. The electron density of states possesses the characteristic asymmetric form and the pseudogap is observed above the critical temperature.       
\end{abstract}
\maketitle
%

\section{Introduction}

Superconductors, in which the condensate of the Cooper pairs is induced by the electron-phonon (EPh) interaction, can be modeled by Fr{\"o}hlich Hamiltonian \cite{Frohlich1950A, Frohlich1952A, Frohlich1954A}. Starting from Fr{\"o}hlich operator, and using the canonical transformation to eliminate the phonon degrees of freedom, the effective Hamiltonian of the BCS theory can be obtained \cite{Frohlich1952A, Bardeen1957A, Bardeen1957B, Fetter1971A}. Then, the equation for the order parameter is derived. It should be noted that, due to the used approximations, the present model provides the quantitative description of the superconducting state only in the weak-coupling limit, so in the fairly narrow group of the low-temperature superconductors \cite{Carbotte1990A, Carbotte2003A}. An important alternative to the above scheme is the approach taken by Eliashberg \cite{Eliashberg1960A}. In the case under consideration, the system of equations determining the thermodynamic properties of the superconducting state is obtained directly from the Fr{\"o}hlich Hamiltonian. Note that the required calculations are performed using the Green matrix functions. Then, the Dyson formula is obtained at the level of the second order equations of motion \cite{Carbotte2003A, Gasser1999A}. 

The Eliashberg equations can be divided into three groups. The first one allows to determine the function of the order parameter ($\varphi_{\kvec}\left(i\omega_{n}\right)$), the second one serves to calculate the value of the wave function renormalization factor ($Z_{\kvec}\left(i\omega_{n}\right)$), and the third one determines the energy shift function ($\chi_{\kvec}\left(i\omega_{n}\right)$). The Eliashberg system is supplemented by the equation for the chemical potential. The quantities ${\bf k}$ and $\omega_{n}$ denote respectively the momentum of the electron and the fermion Matsubara frequency: $\omega_{n}\equiv \left(\pi / \beta\right)\left(2n-1\right)$, where $\beta\equiv\left(k_{B}T\right)^{-1}$ ($k_{B}$ is the Boltzmann constant). It should be noted that the order parameter in the Eliashberg formalism is defined by: $\Delta_{\kvec}\left(i\omega_{n}\right)\equiv \varphi_{\kvec}\left(i\omega_{n}\right)/Z_{\kvec}\left(i\omega_{n}\right)$. The function $Z_{\kvec}\left(i\omega_{n}\right)$ determines the ratio of the electron effective mass to the electron band mass, and the quantity $\chi_{\kvec}\left(i\omega_{n}\right)$ directly renormalizes the electron band energy. The great advantage of the approach proposed by Eliashberg is the fact that the thermodynamic properties of the superconducting state can be determined under this scheme for any value of the electron-phonon coupling constant. The results obtained for the classical superconductors usually provide the quantitative agreement between the theoretical predictions and the experimental data \cite{Carbotte1990A, Szczesniak2007A, Szczesniak2007B, Szczesniak2008A, Szczesniak2010A}. 

The Eliashberg equations can be reduced to the BCS model, but it is necessary to skip the retardation, strong-coupling, and many-body effects: $\Delta_{\kvec}\left(i\omega_{n}\right)\rightarrow \Delta_{\kvec}$, $Z_{\kvec}\left(i\omega_{n}\right)=1$, and $\chi_{\kvec}\left(i\omega_{n}\right)=0$.   

The main purpose of the presented paper is to analyze the properties of the energy gap in the electron density of states in the framework of the Eliashberg formalism. We are interested in the case of the gap, which exists also above the critical temperature ($T_{C}$). Such gap is usually described as the pseudogap in difference of the energy gap induced by the order parameter. The inspiration for the conducted research was the experimental data obtained in the high-temperature superconductors based on cooper \cite{Bednorz1986A, Bednorz1988A, Dagotto1994A}. In the considered group of the materials, the pseudogap is observed, inter alia, in the photoemission spectra \cite{King1995A}, by the tunneling spectroscopy \cite{Renner1998A, Renner1998B, Fischer2007A}, and in the ARPES data \cite{Marshall1996A, Loeser1996A, Ding1996A, Damascelli2003A}. However, we do not set ourselves to quantitative characterize the pseudogap in cuprates due to the fact of neqlecting of the Coulomb electron correlations, which are playing the important role in the considered superconductors \cite{Emery1987A, Littlewood1989A, Hybertsen1990A}.

\section{Eliashberg approach, pseudogap, and pairing mechanism}

We demand that the pseudogap would appear spontaneously in the electron density of states, and not be artificially postulated in the formalism. In order to initially approximate the considered issue let us take into account the isotropic spectral function ($A\left(\omega,\varepsilon\right)$) that uniquely determines the electron density of states: $N\left(\omega\right)\equiv A\left(\omega,0\right)$. The quantity $A\left(\omega,\varepsilon\right)$ should be calculated on the basis of the formula:
$A\left(\omega,\varepsilon\right)=-\frac{1}{\pi}{\rm Im}\left[G_{\rm diag}\left(\omega,\varepsilon\right)\right]$, while the diagonal part of the Green function takes the form:
\begin{equation}
\label{r1(0)}
G_{\rm diag}\left(\omega,\varepsilon\right)=\frac{Z\left(\omega\right)\omega+\varepsilon+\chi\left(\omega\right)}
{\left(Z\left(\omega\right)\omega\right)^{2}-\left(\varepsilon+\chi\left(\omega\right)\right)^{2}-\varphi^{2}\left(\omega\right)}.
\end{equation}
Let us note that on the real axis $Z\left(\omega\right)$, $\chi\left(\omega\right)$, and $\varphi\left(\omega\right)$ take the complex values, thus the formula \eq{r1(0)} can denote a very complicated function. For the case $T>T_{C}$, we assume $\varphi\left(\omega\right)=0$. However, the disappearance of the physical value of the order parameter above the critical temperature ($\Delta\left(T\right)=\varphi\left(T\right)/Z\left(T\right)$), determined with the help of the equation:
\begin{equation}
\label{r2(0)}
\Delta\left(T\right)={\rm Re}\left[\Delta\left(\omega=\Delta\left(T\right),T\right)\right], 
\end{equation}
does not automatically mean the disappearance of the function $\varphi\left(\omega\right)$ on the whole real axis. Mindful of this notice, we write down the simplified formula on the electron density of states:
\begin{equation}
\label{r3(0)}
N\left(\omega\right)\sim
\frac{Z_{I}\left(\omega\right)\omega-\chi_{I}\left(\omega\right)}
{\left(Z_{R}\left(\omega\right)\omega-\varepsilon-\chi_{I}\left(\omega\right)\right)^{2}+\left(Z_{I}\left(\omega\right)\omega-\chi_{I}\left(\omega\right)\right)^{2}}.
\end{equation}
The indexes $R$ and $I$ denote the real and imaginary part, respectively. On the basis of the equation \eq{r3(0)}, we conclude that if the pseudogap in the electron density of states has to be observed above $T_{C}$, is will has to be induced by the strong-coupling and many-body effects 
($Z\left(\omega\right)$ and $\chi\left(\omega\right)$). Hence, the following conclusions can be drawn:\\
- The pseudogap cannot be described in the mean-field approximation, where $Z\left(\omega\right)=1$ and $\chi\left(\omega\right)=0$. It is the dynamic effect, i.e. it is related to the dependence of the wave function renormalization factor and the energy shift function on the frequency. The description of the pseudogap in the static approach is incorrect.\\ 
- The pseudogap has nothing in common with the additional electron phase, because $Z\left(\omega\right)$ and $\chi\left(\omega\right)$ do not represent the order parameter.\\ 
- $Z\left(\omega\right)$ and $\chi\left(\omega\right)$ take the non-zero values both below and above $T_{C}$. Thus, they may make the important contribution  in the energy gap in both superconducting and "normal" state.\\
- Due to the weak dependence of $Z\left(\omega\right)$ and $\chi\left(\omega\right)$ on the temperature, in the case of the significant pseudogap’s existence, we can expect the anomalous temperature dependence of the energy gap in the superconducting state. The energy gap in the superconducting state should smoothly transit into the pseudogap.
As it was already mentioned, our goal is to quite simply describe the properties of the energy gap existing also above $T_{C}$. For this reason, we consider incomplete pairing mechanism associated with the superconducting state in cuprates. We limit to the electron-phonon component, which consists of the electron-phonon and electron-electron-phonon ineraction. This choice was based on the following experimental and theoretical facts:\\
- (i) in cuprates, the ARPES method clearly indicates the existence of the low-energy kink in the energy spectrum near the phonon energy \cite{Damascelli2003A, Cuk2005A}, (ii) with its help one can also observe the isotopic effect associated with the real part of the self-energy \cite{Gweon2004A}. Although, these data are strongly undermined in the paper \cite{Douglas2007A}, (iii) the isotope effect is seen for the critical temperature, especially in the highly undoped area \cite{Franck1994A}, (iv) the direct observations of phonons can be obtained by using the scanning tunneling microscopy \cite{Lee2006A}, and (v) phonons affect the penetration depth and Raman measurements \cite{Hofer2000A}, \cite{Schneider2005A}.
From the theoretical point of view, the idea to use the Hamiltonian that explicitly takes into account the phonon degrees of freedom for the description of the high-temperature superconducting state is not new. The first advanced attempts in this direction were performed by Kim and Tesanovic \cite{Kim1993A}. The authors shown that strong Coulomb correlations may not completely suppress the phonon mechanism in the large doping regime.
\\
- The important role of EEPh interaction in cuprates is evidenced by the experimentally observed half-value of the magnetic flux \cite{Schneider2004A}. It suggests that the electron quartets are formed. The existence of quartets can be the most easily reproduced with the help of the EEPh interaction 
- for this purpose is sufficient to use the canonical transformation that eliminates the phonon degrees of freedom \cite{Szczesniak2012G, Szczesniak2011D}.
It is worth to emphasize that ability to complete the BCS theory with the four-fermion interaction was first time noted by Rickayzen \cite{Rickayzen1964A}, who motivated his idea as analogous to nuclear physics, where alpha particle represents the stable system of four fermions. In the context of the description of the superconducting state, Rickayzen's idea was followed after many years in \cite{Mackowiak1998A, Mackowiak2000A, Mackowiak2011A}.
\\
- Carried out so far theoretical studies suggest that using the Hamiltonian including EPh and EEPh terms it is possible to reproduce some properties of the superconducting state in cuprates. In particular, it was shown that the model allows to correctly calculate the critical temperature and the energy gap at zero Kelvin. It should be noted that appropriate calculations were carried out for the large number of the chemical compounds, and each time obtained results were consistent with the experimental data \cite{Szczesniak2012G, Szczesniak2011D, Szczesniak2012H, Szczesniak2014C, Szczesniak2014D, Szczesniak2014E, Szczesniak2014H, Szczesniak2015C}.

\vspace{-1cm}
\section{The formalism}
The Hamiltonian, which models the electron-phonon and electron-electron-phonon interaction, in momentum representation has the following form \cite{Szczesniak2012G, Szczesniak2011D}:
\begin{equation}
\label{r1(1.0)}
H\equiv H^{\left(0\right)}+H^{\left(1\right)}+H^{\left(2\right)},
\end{equation}
where first term describes the system of non-interacting electrons and phonons:
\begin{equation}
\label{r2(1.0)}
H^{\left(0\right)}\equiv\sum_{\kvec\sigma }\overline\varepsilon _{\kvec}c_{\kvec\sigma
}^{\dagger}c_{\kvec\sigma }+\sum_{\qvec}\omega _{\qvec}b_{\qvec}^{\dagger}b_{\qvec}.
\end{equation}
The symbol $\overline\varepsilon _{\kvec}$ is defined by: $\overline\varepsilon _{\kvec}\equiv \varepsilon_{\kvec}-\mu$, whereas $\varepsilon _{\kvec}$ is the electron band energy; $\mu$ denotes the chemical potential. We assume that electrons can propagate in the square lattice with the hopping integral $t$. In the considered case, energy band can be calculated using: 
$\varepsilon _{\kvec}=-t\gamma\left({\bf k}\right)$, where $\gamma\left({\bf k}\right)\equiv 2\left[\cos\left(k_{x}\right)+\cos\left(k_{y}\right)\right]$. The function $\omega_{\qvec}$ determines the values of the phonon energies.
The interaction operators are given by:
\begin{equation}
\label{r3(1.0)}
H^{\left(1\right)}\equiv\sum_{\kvec\qvec\sigma }v^{\left(1\right)}_{\kvec}\left({\bf q}\right)
c_{\kvec+\qvec\sigma}^{\dagger}c_{\kvec\sigma}\phi_{\qvec},
\end{equation}
and
\begin{equation}
\label{r4(1.0)}
H^{\left(2\right)}\equiv\sum_{\kvec\kvec^{'}\qvec\lvec\sigma}
v^{\left(2\right)}_{\kvec,\kvec^{'}}\left({\bf q},{\bf l}\right)
c_{\kvec-\lvec\sigma }^{\dagger}c_{\kvec\sigma}
c_{\kvec^{'}+\lvec+\qvec-\sigma}^{\dagger}c_{\kvec^{'}-\sigma}\phi_{\qvec},
\end{equation}
where: $\phi_{\qvec}\equiv b_{-\qvec}^{\dagger}+b_{\qvec}$.
The quantities $v^{\left(1\right)}_{\kvec}\left({\bf q}\right)$ and $v^{\left(2\right)}_{\kvec,\kvec^{'}}\left({\bf q},{\bf l}\right)$ denote the EPh and EEPh potentials, respectively. Next, we define the Nambu spinors: 
\begin{equation}
\label{r5(1.0)}
\Psi _{\kvec}\equiv\left( 
\begin{array}{l}
c_{\kvec\uparrow } \\ 
c_{-\kvec\downarrow}^{\dagger}
\end{array}
\right) ,\Psi _{\kvec}^{\dagger}\equiv\left( 
\begin{array}{ll}
c_{\kvec\uparrow }^{\dagger} & c_{-\kvec\downarrow}
\end{array}
\right).
\end{equation}
The matrix Green function $G_{\kvec}(i\omega _n)\equiv\left<\left<\Psi_{\kvec}|\Psi^{\dagger}_{\kvec}\right>\right>_{i\omega_{n}}$ takes the form: 
\begin{eqnarray}
\label{r6(1.0)} 
& &\left( \begin{array}{cc}
<< c_{\kvec\uparrow}|c_{\kvec\uparrow}^{\dagger}>>_{i\omega_{n}}&
<< c_{\kvec\uparrow}|c_{-\kvec\downarrow}>>_{i\omega_{n}} \\
<< c_{-\kvec\downarrow}^{\dagger}|c_{\kvec\uparrow}^{\dagger}>>_{i\omega_{n}}&
<< c_{-\kvec\downarrow}^{\dagger}|c_{-\kvec\downarrow}>>_{i\omega_{n}}
\end{array}\right)\\ \nonumber
&\equiv&\left( \begin{array}{cc}
g^{\left(A\right)}_{\kvec}\left({i\omega_{n}}\right)&
g^{\left(B\right)}_{\kvec}\left({i\omega_{n}}\right) \\
g^{\left(C\right)}_{\kvec}\left({i\omega_{n}}\right)&
g^{\left(D\right)}_{\kvec}\left({i\omega_{n}}\right)
\end{array}\right).
\end{eqnarray}
For $T>T_{C}$, the diagonal elements of $G_{\kvec}(i\omega _n)$ model the properties of the non-superconducting state. The condensate of Cooper pairs below the critical temperature is described by all elements of the matrix Green function. The Nambu spinors do not meet strictly anti-commutation relation \cite{Szczesniak2005A}. Hence, the elements of $G_{\kvec}(i\omega _n)$ should be calculated directly from the anti-commutation relation for the fermion operators. 
The equations for the scalar Green functions have a following form: 

\begin{equation}
\label{r1(2.0)}  
g^{\left(A\right)}_{\kvec}\left(i\omega_{n}\right)=g^{\left(A\right)}_{0\kvec}\left(i\omega_{n}\right)+
g^{\left(A\right)}_{0\kvec}\left(i\omega_{n}\right)m^{\left(A\right)}_{\kvec}\left(i\omega_{n}\right)
g^{\left(A\right)}_{0\kvec}\left(i\omega_{n}\right),
\end{equation}

\begin{equation}
\label{r2(2.0)}  
g^{\left(B\right)}_{\kvec}\left(i\omega_{n}\right)=
g^{\left(A\right)}_{0\kvec}\left(i\omega_{n}\right)m^{\left(B\right)}_{\kvec}\left(i\omega_{n}\right)
g^{\left(B\right)}_{0\kvec}\left(i\omega_{n}\right),
\end{equation}

\begin{equation}
\label{r3(2.0)}  
g^{\left(C\right)}_{\kvec}\left(i\omega_{n}\right)=
g^{\left(B\right)}_{0\kvec}\left(i\omega_{n}\right)m^{\left(C\right)}_{\kvec}\left(i\omega_{n}\right)
g^{\left(A\right)}_{0\kvec}\left(i\omega_{n}\right),
\end{equation}

\begin{equation}
\label{r4(2.0)}  
g^{\left(D\right)}_{\kvec}\left(i\omega_{n}\right)=g^{\left(B\right)}_{0\kvec}\left(i\omega_{n}\right)+
g^{\left(B\right)}_{0\kvec}\left(i\omega_{n}\right)m^{\left(D\right)}_{\kvec}\left(i\omega_{n}\right)
g^{\left(B\right)}_{0\kvec}\left(i\omega_{n}\right),
\end{equation}
where the non-interacting Green functions are defined as: 
\begin{equation}
\label{r5(2.0)}  
g^{\left(A\right)}_{0\kvec}\left(i\omega_{n}\right)\equiv\frac{1}{i\omega_{n}-\overline\varepsilon_{\kvec}},\hspace{2mm}
{\rm and} \hspace{2mm}
g^{\left(B\right)}_{0\kvec}\left(i\omega_{n}\right)\equiv\frac{1}{i\omega_{n}+\overline\varepsilon_{\kvec}}. 
\end{equation}
The scalar Dyson equations were achieved only for diagonal Green function. The non-diagonal elements describing the properties of the superconducting state meet the truncated Dyson equation due to the existence of the condensate only in the case of non-zero pairing interaction. The self-energies $m^{\left(A\right)}_{\kvec}\left(i\omega_{n}\right)$-$m^{\left(D\right)}_{\kvec}\left(i\omega_{n}\right)$ can be written as:
\begin{widetext}
\begin{equation}
\label{r7(2.0)}  
m^{\left(A\right)}_{\kvec}\left(i\omega_{n}\right)=-\left(v_{1}+v_{2}\left<n\right>\right)^{2}\frac{1}{\beta}\sum_{\omega_{m}}
\sum_{\qvec}<< c _{\kvec-\qvec\uparrow}| c _{\kvec-\qvec\uparrow}^{\dagger}>>_{i\omega_{m}}<<\phi _{\qvec}|\phi _{-\qvec}>>_{i\left(\omega_{n}-\omega_{m}\right)},
\end{equation}
\begin{eqnarray}
\label{r8(2.0)}  
m^{\left(B\right)}_{\kvec}\left(i\omega_{n}\right)&=&\left(v_{1}^{2}+2v_{1}v_{2}\left<n\right>\right)\frac{1}{\beta}\sum_{\omega_{m}}
\sum_{\qvec}<< c _{\kvec-\qvec\uparrow}| c _{-\kvec+\qvec\downarrow}>>_{i\omega_{m}}<<\phi _{\qvec}|\phi _{-\qvec}>>_{i\left(\omega_{n}-\omega_{m}\right)}
\\ \nonumber
&-&\frac{4v_{2}^{2}}{\beta^{3}}\sum_{\omega_{m_{1}}\sim\omega_{m_{3}}}
\left[\frac{1}{N}\sum_{\kvec_{1}}<< c _{\kvec_{1}\downarrow}| c _{-\kvec_{1}\uparrow}>>_{i\omega_{m_{1}}}\right]
\left[\frac{1}{N}\sum_{\kvec_{2}}<< c _{-\kvec_{2}\downarrow}^{\dagger}|c_{\kvec_{2}\uparrow}^{\dagger}>>_{i\omega_{m_{2}}}\right]
\\ \nonumber
&\times&\sum_{\qvec}<< c _{\kvec-\qvec\uparrow}| c _{-\kvec+\qvec\downarrow}>>_{i\omega_{m_{3}}}<<\phi _{\qvec}|\phi _{-\qvec}>>_{i\left(\omega_{n}-\omega_{m_{1}}-\omega_{m_{2}}-\omega_{m_{3}}\right)},
\end{eqnarray}
\begin{eqnarray}
\label{r9(2.0)}  
m^{\left(C\right)}_{\kvec}\left(i\omega_{n}\right)&=&\left(v_{1}^{2}+2v_{1}v_{2}\left<n\right>\right)\frac{1}{\beta}\sum_{\omega_{m}}
\sum_{\qvec}<<c_{-\kvec+\qvec\downarrow}^{\dagger}|c_{\kvec-\qvec\uparrow}^{\dagger}>>_{i\omega_{m}}
<<\phi _{\qvec}|\phi _{-\qvec}>>_{i\left(\omega_{n}-\omega_{m}\right)}
\\ \nonumber
&-&\frac{4v_{2}^{2}}{\beta^{3}}\sum_{\omega_{m_{1}}\sim\omega_{m_{3}}}
\left[\frac{1}{N}\sum_{\kvec_{1}}<<c_{-\kvec_{1}\uparrow}^{\dagger}|c_{\kvec_{1}\downarrow}^{\dagger}>>_{i\omega_{m_{1}}}\right]
\left[\frac{1}{N}\sum_{\kvec_{2}}<< c_{\kvec_{2}\uparrow}|c_{-\kvec_{2}\downarrow}>>_{i\omega_{m_{2}}}\right]
\\ \nonumber
&\times&\sum_{\qvec}<< c_{-\kvec+\qvec\downarrow}^{\dagger}|c_{\kvec-\qvec\uparrow}^{\dagger}>>_{i\omega_{m_{3}}}<<\phi _{\qvec}|\phi _{-\qvec}>>_{i\left(\omega_{n}-\omega_{m_{1}}-\omega_{m_{2}}-\omega_{m_{3}}\right)},
\end{eqnarray}
\begin{equation}
\label{r10(2.0)}  
m^{\left(D\right)}_{\kvec}\left(i\omega_{n}\right)=-\left(v_{1}+v_{2}\left<n\right>\right)^{2}\frac{1}{\beta}\sum_{\omega_{m}}\sum_{\qvec}
<< c^{\dagger}_{-\kvec+\qvec\downarrow}|c_{-\kvec+\qvec\downarrow}>>_{i\omega_{m}}
<<\phi _{\qvec}|\phi _{-\qvec}>>_{i\left(\omega_{n}-\omega_{m}\right)}.
\end{equation}
\end{widetext}
New symbols occurring in the formulas \eq{r7(2.0)}-\eq{r10(2.0)} were defined in the following way: 
$v^{\left(1\right)}_{\kvec}\left(\bf{q}\right)\rightarrow v_{1}$, 
$v^{\left(2\right)}_{\kvec_{1},\kvec_{2}}\left({\bf l}_{1},{\bf l}_{2}\right)\rightarrow \frac{v_{2}}{N}$, and
$\left<n\right>\equiv\frac{1}{N}\sum_{\kvec\sigma}\left<c^{\dagger}_{\kvec\sigma}c_{\kvec\sigma}\right>$.
The phonon propagator can be calculated in the non-interacting case: 
\begin{eqnarray}
\label{r11(2.0)}  
<<\phi _{\qvec}|\phi _{-\qvec}>>_{i\omega_{l}}&=&
-2\int_{0}^{\omega_{0}}d{\Omega}F_{\qvec}\left(\Omega \right)\frac{\Omega}{\omega_{l}^2+\Omega^{2}}\\ \nonumber
&\equiv& -P_{\qvec}\left(i\omega_{l}\right),                      
\end{eqnarray}
where: $\omega_{l}\equiv \left(2\pi/\beta\right)l$, $F_{\qvec}\left(\Omega \right)$ represents the phonon density of states, and $\omega_{0}$ is the maximum phonon frequency. Obtained results enable the derivation of the Dyson equation for the matrix Green function: 
\begin{equation}
\label{r1(3.0)} 
G_{\kvec}(i\omega _n)=G_{0\kvec}(i\omega _n)+G_{0\kvec}(i\omega _n)M_{\kvec}(i\omega _n)G_{0\kvec}(i\omega _n),
\end{equation}
where $G_{0\kvec}(i\omega_n)\equiv\left( i\omega_n\tau_0-\overline\varepsilon_{\kvec}\tau_{3}\right)^{-1}$ denotes the propagator for the non-interacting electrons.
The matrix of self-energy has the form: 
\begin{widetext}
\begin{eqnarray}
\label{r3(3.0)}
M_{\kvec}(i\omega _n)&=& 
\left(v_{1}+v_{2}\left<n\right>\right)^{2}\frac{1}{\beta}\sum_{\omega_{m}\qvec}\tau_{3}G_{\kvec-\qvec}\left(i\omega_{m}\right)\tau_{3}
P_{\qvec}\left(i\omega_{n}-i\omega_{m}\right)\\ \nonumber
&+&\frac{v_{2}^{2}\left<n\right>^{2}}{4}\frac{1}{\beta}\sum_{\omega_{m}\qvec}\left(\tau_{0}-\tau_{3}\right)G_{\kvec-\qvec}\left(i\omega_{m}\right)
\left(\tau_{0}+\tau_{3}\right)P_{\qvec}\left(i\omega_{n}-i\omega_{m}\right)\\ \nonumber
&+&\frac{v_{2}^{2}\left<n\right>^{2}}{4}\frac{1}{\beta}\sum_{\omega_{m}\qvec}\left(\tau_{0}+\tau_{3}\right)G_{\kvec-\qvec}\left(i\omega_{m}\right)
\left(\tau_{0}-\tau_{3}\right)P_{\qvec}\left(i\omega_{n}-i\omega_{m}\right)\\ \nonumber
&-& \frac{v_{2}^{2}}{\beta^{3}}\sum_{\omega_{m_{1}}\sim\omega_{m_{3}}}
\frac{1}{N}\sum_{\kvec_{1}}\left[G^{\star}_{\kvec_{1}}\left(i\omega_{m_{1}}\right)\right]_{12}
\frac{1}{N}\sum_{\kvec_{2}}\left[G_{\kvec_{2}}\left(i\omega_{m_{2}}\right)\right]_{12}\\ \nonumber
&\times&\sum_{\qvec}\left(\tau_{0}-\tau_{3}\right)G_{\kvec-\qvec}\left(i\omega_{m_{3}}\right)\left(\tau_{0}+\tau_{3}\right)
P_{\qvec}\left(i\omega_{n}-i\omega_{m_{1}}-i\omega_{m_{2}}-i\omega_{m_{3}}\right)\\ \nonumber
&-& \frac{v_{2}^{2}}{\beta^{3}}\sum_{\omega_{m_{1}}\sim\omega_{m_{3}}}
\frac{1}{N}\sum_{\kvec_{1}}\left[G^{\star}_{\kvec_{1}}\left(i\omega_{m_{1}}\right)\right]_{21}
\frac{1}{N}\sum_{\kvec_{2}}\left[G_{\kvec_{2}}\left(i\omega_{m_{2}}\right)\right]_{21}\\ \nonumber
&\times&\sum_{\qvec}\left(\tau_{0}+\tau_{3}\right)G_{\kvec-\qvec}\left(i\omega_{m_{3}}\right)\left(\tau_{0}-\tau_{3}\right)
P_{\qvec}\left(i\omega_{n}-i\omega_{m_{1}}-i\omega_{m_{2}}-i\omega_{m_{3}}\right),
\end{eqnarray}
\end{widetext}
where $\tau_{0}$-$\tau_{3}$ represent the Pauli matrix base and the following dependence was used: $\left<\left<A|B^{\dagger}\right>\right>_{i\omega_{n}}=-\left<\left<A^{\dagger}|B\right>\right>_{i\omega_{n}}^{\star}$. The symbol $^{\star}$ is the complex coupling. Next, the distribution of self-energy was postulated: 
\begin{eqnarray}
\label{r5(3.0)}
M_{\kvec}\left(i\omega_{n}\right)&\equiv & i\omega_{n}\left(1-Z_{\kvec}\left(i\omega_{n}\right)\right)\tau_{0}
+\chi_{\kvec}\left(i\omega_{n}\right)\tau_{3}\\ \nonumber
&+&\varphi_{\kvec}\left(i\omega_{n}\right)\tau_{1}+\overline\varphi_{\kvec}\left(i\omega_{n}\right)\tau_{2}.
\end{eqnarray}
On the other hand, the equation \eq{r1(3.0)} can be written as:
\begin{equation}
\label{r6(3.0)}
G_{\kvec}^{-1}\left(i\omega_{n}\right)=G_{0\kvec}^{-1}\left(i\omega_{n}\right)-M_{\kvec}\left(i\omega_{n}\right).
\end{equation}
Substituting the expression \eq{r5(3.0)} into \eq{r6(3.0)} and calculating the overt form of the Green function, it was obtained:
\begin{widetext}
\begin{eqnarray}
\label{r7(3.0)}
G_{\kvec}\left(i\omega_{n}\right)=
-D_{\kvec}^{-1}\left(i\omega_{n}\right)
\left[i\omega_{n}Z_{\kvec}\left(i\omega_{n}\right)\tau_{0}+\left(\chi_{\kvec}\left(i\omega_{n}\right)+\overline\varepsilon_{\kvec}\right)\tau_{3}+\varphi_{\kvec}\left(i\omega_{n}\right)\tau_{1}+\overline\varphi_{\kvec}\left(i\omega_{n}\right)\tau_{2}\right], 
\end{eqnarray}
\end{widetext}
where:  
$D_{\kvec}\left(i\omega_{n}\right)\equiv \left(\omega_{n}Z_{\kvec}\left(i\omega_{n}\right)\right)^{2}+
\left(\chi_{\kvec}\left(i\omega_{n}\right)+\overline\varepsilon_{\kvec}\right)^{2}+\varphi^{2}_{\kvec}\left(i\omega_{n}\right)
+\overline\varphi^{2}_{\kvec}\left(i\omega_{n}\right)$. In the next step, the formula \eq{r7(3.0)} should be inserted into \eq{r3(3.0)}:  
\begin{widetext}
\begin{eqnarray}
\label{r9(3.0)}
M_{\kvec}\left(i\omega_{n}\right)=
&-&i\left(v_{1}+v_{2}\left<n\right>\right)^{2}\frac{1}{\beta}\sum_{\omega_{m}\qvec}\omega_{m}Z_{\kvec-\qvec}\left(i\omega_{m}\right)
D^{-1}_{\kvec-\qvec}\left(i\omega_{m}\right)P_{\qvec}\left(i\omega_{n}-i\omega_{m}\right)\tau_{0}\\ \nonumber
&-&
\left(v_{1}+v_{2}\left<n\right>\right)^{2}\frac{1}{\beta}\sum_{\omega_{m}\qvec}\left(\chi_{\kvec-\qvec}\left(i\omega_{m}\right)+
\overline\varepsilon_{\kvec-\qvec}\right)
D^{-1}_{\kvec-\qvec}\left(i\omega_{m}\right)P_{\qvec}\left(i\omega_{n}-i\omega_{m}\right)\tau_{3}\\ \nonumber
&+&
\left(v_{1}^{2}+2v_{1}v_{2}\left<n\right>\right)\frac{1}{\beta}\sum_{\omega_{m}\qvec}\varphi_{\kvec-\qvec}\left(i\omega_{m}\right)
D^{-1}_{\kvec-\qvec}\left(i\omega_{m}\right)P_{\qvec}\left(i\omega_{n}-i\omega_{m}\right)\tau_{1}\\ \nonumber
&+&
\frac{4v_{2}^{2}}{\beta^{3}}\sum_{\omega_{m_{1}}\sim\omega_{m_{3}}}
\frac{1}{N}\sum_{\kvec_{1}}\left[\varphi_{\kvec_{1}}\left(i\omega_{m_{1}}\right)-i
\overline\varphi_{\kvec_{1}}\left(i\omega_{m_{1}}\right)\right]
D^{-1}_{\kvec_{1}}\left(i\omega_{m_{1}}\right)\\ \nonumber
&\times&\frac{1}{N}\sum_{\kvec_{2}}\left[\varphi_{\kvec_{2}}\left(i\omega_{m_{2}}\right)+i
\overline\varphi_{\kvec_{2}}\left(i\omega_{m_{2}}\right)\right]
D^{-1}_{\kvec_{2}}\left(i\omega_{m_{2}}\right)\\ \nonumber
&\times&
\sum_{\qvec}\varphi_{\kvec-\qvec}\left(i\omega_{m_{3}}\right)D^{-1}_{\kvec-\qvec}\left(i\omega_{m_{3}}\right)
P_{\qvec}\left(i\omega_{n}-i\omega_{m_{1}}-i\omega_{m_{2}}-i\omega_{m_{3}}\right)\tau_{1}\\ \nonumber
&+&
\left(v_{1}^{2}+2v_{1}v_{2}\left<n\right>\right)\frac{1}{\beta}\sum_{\omega_{m}\qvec}\overline\varphi_{\kvec-\qvec}\left(i\omega_{m}\right)
D^{-1}_{\kvec-\qvec}\left(i\omega_{m}\right)P_{\qvec}\left(i\omega_{n}-i\omega_{m}\right)\tau_{2}\\ \nonumber
&+&
\frac{4v_{2}^{2}}{\beta^{3}}\sum_{\omega_{m_{1}}\sim\omega_{m_{3}}}
\frac{1}{N}\sum_{\kvec_{1}}\left[\varphi_{\kvec_{1}}\left(i\omega_{m_{1}}\right)-i
\overline\varphi_{\kvec_{1}}\left(i\omega_{m_{1}}\right)\right]
D^{-1}_{\kvec_{1}}\left(i\omega_{m_{1}}\right)\\ \nonumber
&\times&\frac{1}{N}\sum_{\kvec_{2}}\left[\varphi_{\kvec_{2}}\left(i\omega_{m_{2}}\right)+i
\overline\varphi_{\kvec_{2}}\left(i\omega_{m_{2}}\right)\right]
D^{-1}_{\kvec_{2}}\left(i\omega_{m_{2}}\right)\\ \nonumber
&\times&
\sum_{\qvec}\overline\varphi_{\kvec-\qvec}\left(i\omega_{m_{3}}\right)D^{-1}_{\kvec-\qvec}\left(i\omega_{m_{3}}\right)
P_{\qvec}\left(i\omega_{n}-i\omega_{m_{1}}-i\omega_{m_{2}}-i\omega_{m_{3}}\right)\tau_{2}.\\ \nonumber
\end{eqnarray}
\end{widetext}
Comparing with each other the formulas \eq{r5(3.0)} and \eq{r9(3.0)}, we get the Eliashberg equations. Due to the numerical complexity, the isotropic equations were taken into account, and we assume $v_{2}\rightarrow v_{2}/2$:
\begin{widetext}
\begin{equation}
\label{r1(5.0)}
Z_{n}=1+\left(v_{1}+\frac{v_{2}}{2}\left<n\right>\right)^{2}\frac{1}{\beta\omega_{n}}\sum_{m}P\left(n-m\right)\omega_{m}Z_{m}
\frac{1}{N}\sum_{\qvec}D_{\qvec}^{-1}\left(m\right),
\end{equation}
\begin{equation}
\label{r2(5.0)}
\chi_{n}=-\left(v_{1}+\frac{v_{2}}{2}\left<n\right>\right)^{2}\frac{1}{\beta}\sum_{m}P\left(n-m\right)
\frac{1}{N}\sum_{\qvec}D_{\qvec}^{-1}\left(m\right)\left[\varepsilon_{\qvec}+\chi_{m}-\mu\right],
\end{equation}
\begin{eqnarray}
\label{r3(5.0)}
\varphi_{n}&=&
\left(v_{1}^{2}+v_{1}v_{2}\left<n\right>\right)\frac{1}{\beta}\sum_{m}P\left(n-m\right)\varphi_{m}
\frac{1}{N}\sum_{\qvec}D^{-1}_{\qvec}\left(m\right)\\ \nonumber
&+&
\frac{v_{2}^{2}}{\beta^{3}}\sum_{m_{1}\sim m_{3}}
P\left(n-m_{1}-m_{2}-m_{3}+1\right)
\varphi_{m_{1}}\varphi_{m_{2}}\varphi_{m_{3}}\\ \nonumber
&\times&
\frac{1}{N^{3}}\sum_{\qvec_{1}}D^{-1}_{\qvec_{1}}\left(m_{1}\right)
\sum_{\qvec_{2}}D^{-1}_{\qvec_{2}}\left(m_{2}\right)
\sum_{\qvec_{3}}D^{-1}_{\qvec_{3}}\left(m_{3}\right),
\end{eqnarray}
\end{widetext}
where: $Z_{n}\equiv Z\left(i\omega_{n}\right)$, $\chi_{n}\equiv\chi\left(i\omega_{n}\right)$, $\varphi_{n}\equiv\varphi\left(i\omega_{n}\right)$, and $D_{\qvec}\left(m\right)\equiv D_{\qvec}\left(i\omega_{m}\right)$. It should be noted that $\varphi_{n}$ and $\overline\varphi_{n}$ take identical values. For this reason, the discussion is limited to $\varphi_{n}$. The function $P\left(n\right)$ is given by:
$P\left(n\right)\equiv\frac{\nu^2}{n^2+\nu^2}$, and $\nu\equiv\beta\omega_{0}/2\pi$. The chemical potential should be calculated from the equation: $\left<n\right>=1-\int_{-W}^Wd\varepsilon\rho \left( \varepsilon\right)\tanh\left(\frac{\beta\left(\varepsilon-\mu\right)}{2}\right)$, $W$ is the half-width of the electron band, and the density of states has the form \cite{Szczesniak2002A}: $\rho\left(\varepsilon\right)=b_{1}\ln|\frac{\varepsilon}{b_{2}}|$, where $b_{1}=-0.04687t^{-1}$ and $b_{2}=21.17796t$. From the physical point of view, the function $\rho\left(\varepsilon\right)$ characterizes the van Hove singularity located at the Fermi level. In the cuprates it is related to the ${\rm CuO_{2}}$ planes and is observed close to the Fermi energy. For example, in ${\rm YBa_{2}Cu_{4}O_{8}}$ the singularity exists about $19$ meV below $\varepsilon_{F}$ at the $Y$ point in the Brillouin zone \cite{Gofron1994A}.
  
\section{The analysis of the superconducting and the anomalous normal state}
\begin{figure*}
\includegraphics[scale=0.60]{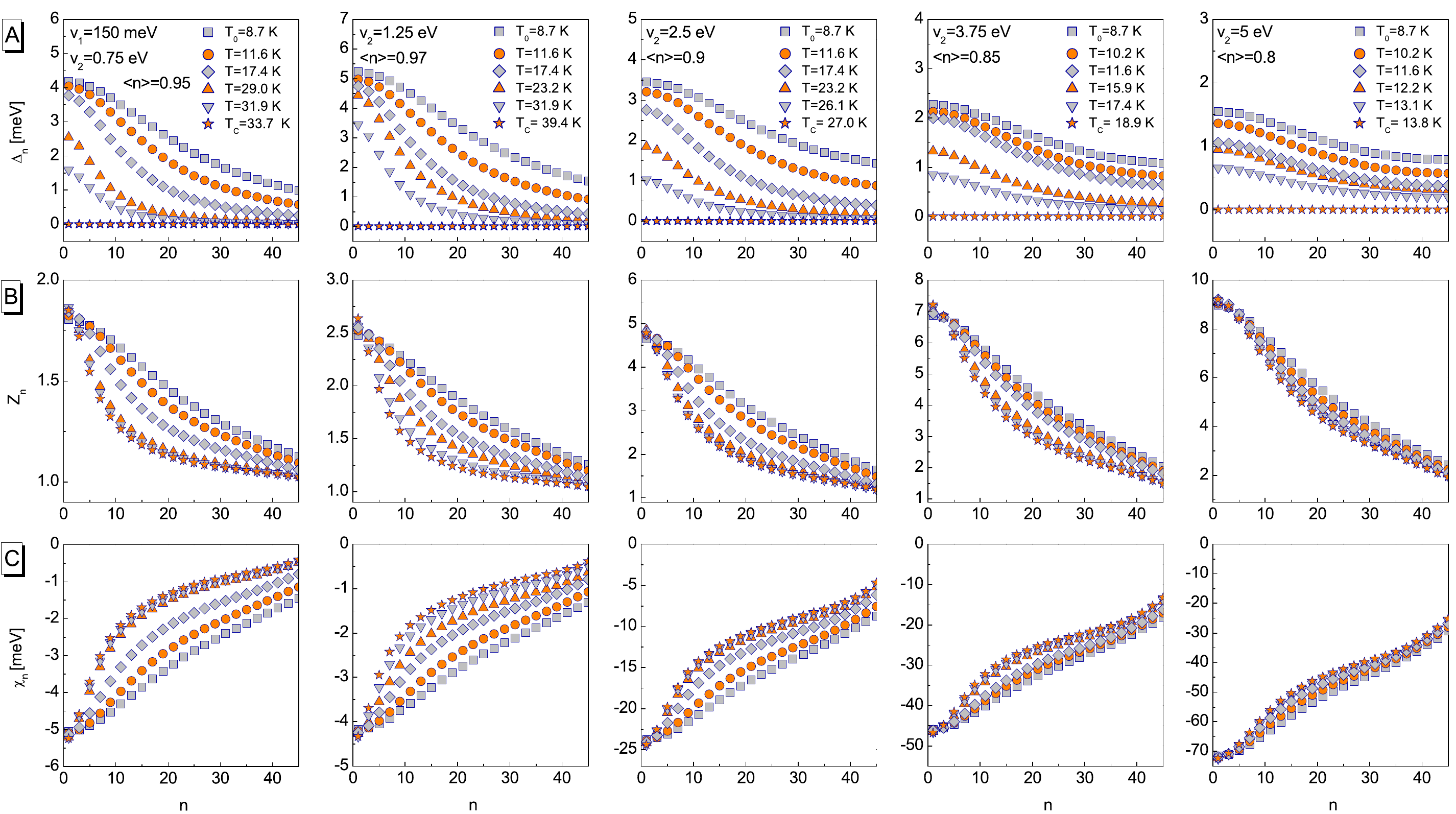}
\caption{
(A) The order parameter, (B) the wave function renormalization factor, and (C) the energy shift function on imaginary axis. 
}
\label{f07}
\end{figure*}

\begin{figure*}
\includegraphics[scale=0.60]{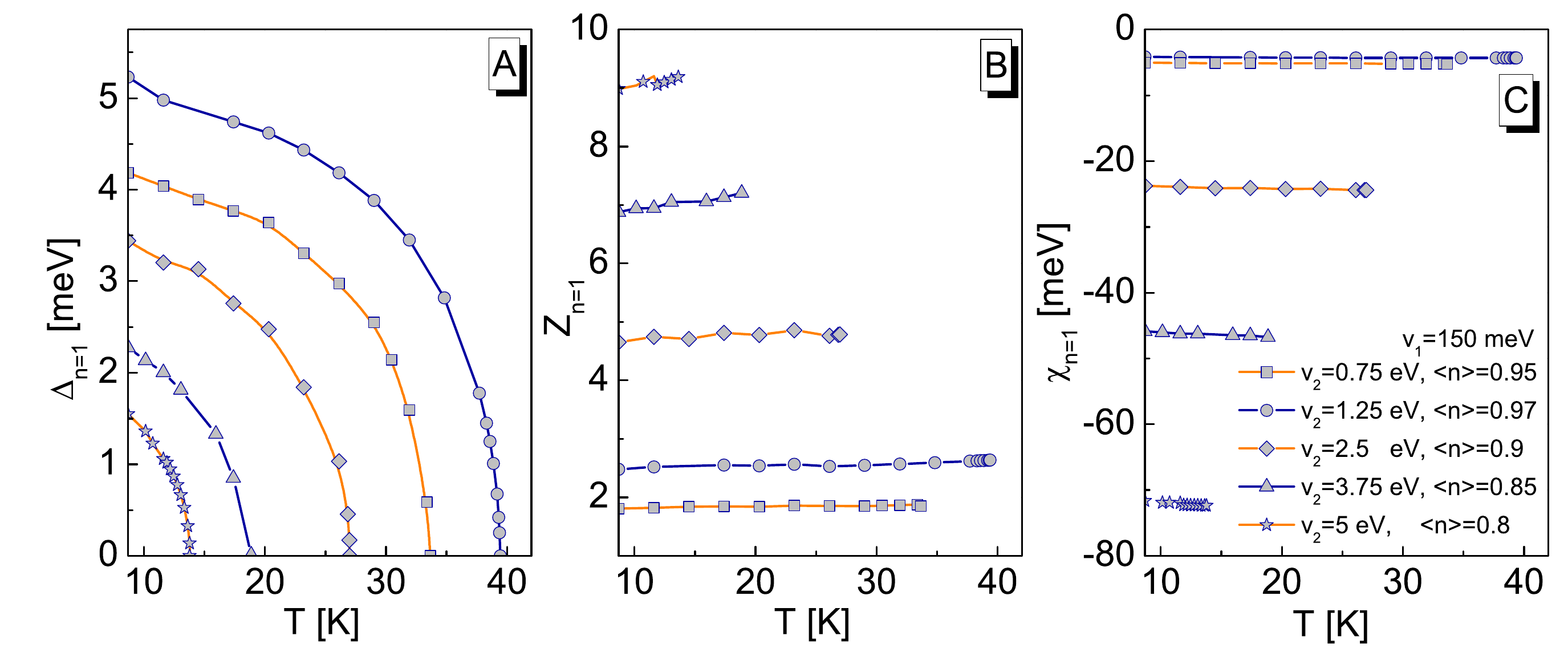}
\caption{
(A) The maximum value of the order parameter, (B) the wave function renormalization factor, and (C) the energy shift function as a function of the temperature.}
\label{f08}
\end{figure*}

The Eliashberg equations set was solved for $45$ Matsubara frequencies, which was related to taking into account above $90$ thousands of terms in the triple sum over Matsubara frequencies. The calculations used the modified numerical methods originally designed to solve the classical Eliashberg equations with the electron-phonon pairing kernel \cite{Szczesniak2006B}. The solutions of the Eliashberg equations are symmetrical: $Z_{m}=Z_{-m+1}$, $\chi_{m}=\chi_{-m+1}$, and $\varphi_{m}=\varphi_{-m+1}$. This property greatly simplifies the numerical calculations. The functions $Z_{m}$, $\chi_{m}$, and $\varphi_{m}$ are stable for the temperatures above $T_{0}=8.7$ K. Results obtained for the superconducting state ($T\in\left<T_{0},T_{C}\right>$) are collected in the figures \fig{f07} (A)-(C). In all cases it has been adopted $v_{1}=150$ meV and the increasing values of the potential $v_{2}$ from $0.75$ eV to $5$ eV were taken into account. The average number of the electrons was chosen in such a way that the order parameter would take its maximum value for $T=T_{0}$. It can be noted that obtained functions have characteristic lorenzian shape, formally similar to the solutions obtained with the help of the classical Eliashberg equations \cite{Szczesniak2006B, Szczesniak2006A}.
%
%
The temperature dependence of the Eliashberg equations solutions can be most conveniently traced by analyzing the form of $\Delta_{n=1}\left(T\right)$, $Z_{n=1}\left(T\right)$, and $\chi_{n=1}\left(T\right)$. In the Eliashberg formalism, $\Delta_{n=1}$ with good approximation reproduces the physical value of the order parameter, $Z_{n=1}$ correctly determines the ratio of the electron effective mass to the electron band mass, and $\chi_{n=1}$ well defines the direct renormalization of the electron band energy. The results are presented in the figure \fig{f08}. Typical curves characterizing the phase transition of the second type were obtained each time (Fig. \fig{f08} (A)). This means that the complete description of the disappearance of the superconducting state requires the precise consideration of the retardation effects (the explicit dependence of the Eliashberg equations solutions on $\omega_{n}$), the strong-coupling effects ($Z$), and the many-body effects ($\chi$). For case when they are omitted (as it was done for the {\it toy} model \cite{Szczesniak2012G, Szczesniak2011D}), appears the non-physical phase transition of the first order. On the other hand, the wave function renormalization factor and the energy shift function weakly depend on the temperature (Figs. \fig{f08} (B) and (C)), which is the typically expected behavior of the considered quantities. It is also necessary to mention that growth of the EEPh interaction leads to the strong renormalization of the electron band mass and the band energy, which is caused by the high values of $Z_{n=1}$ and $\chi_{n=1}$.
%
%

\begin{figure*}
\includegraphics[scale=0.60]{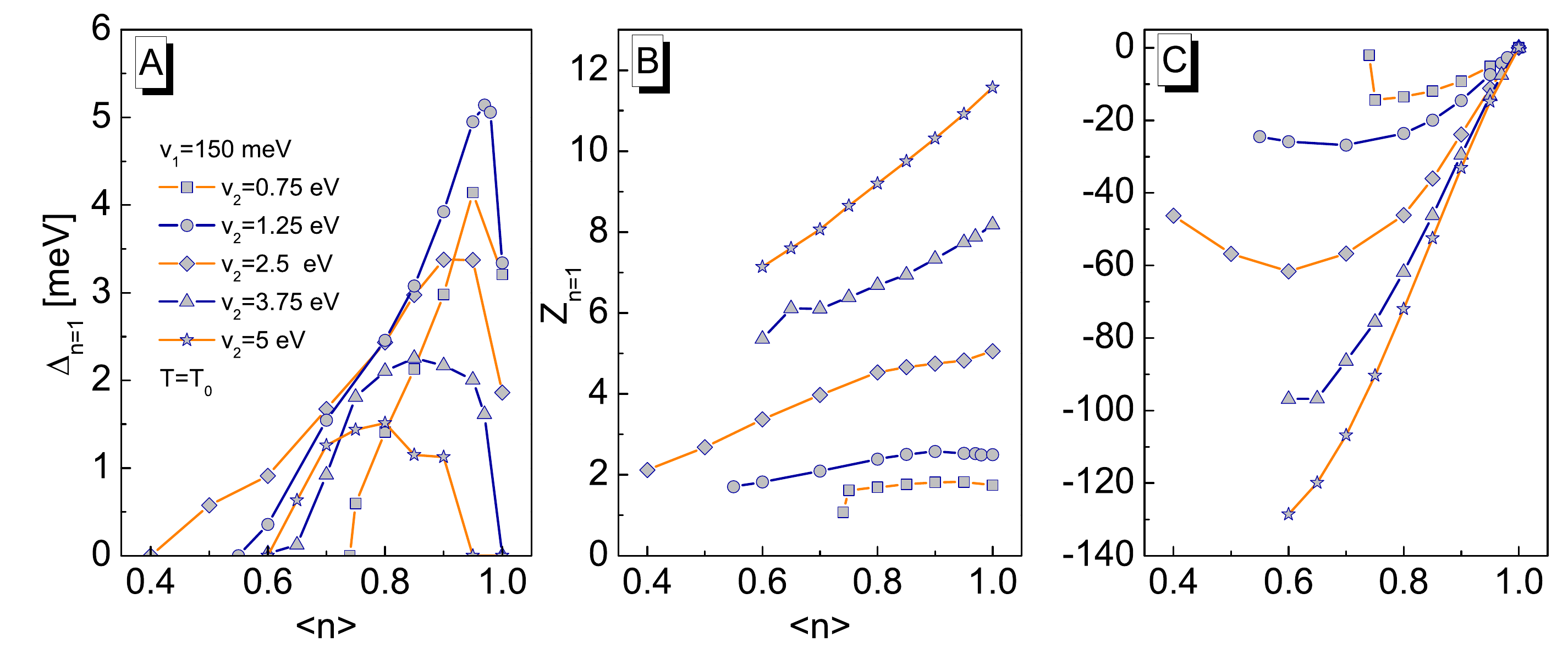}
\caption{
(A) The maximum value of order parameter, (B) the wave function renormalization factor, and (C) the energy shift function as a function of doping. 
}
\label{f09}
\end{figure*}

The figures \fig{f09} (A)-(C) show plots of $\Delta_{n=1}$, $Z_{n=1}$, and $\chi_{n=1}$ in the dependence of the average number of electrons (doping). The most interesting results were obtained for the order parameter (Fig. \fig{f09} (A)). Our model predicts that the dependence of $\Delta_{n=1}$ on $\left<n\right>$ for the low values of the EEPh potential ($v_{2}=0.75$ eV) looks very much like the dependence for the classical superconducting state induced by the electron-phonon interaction \cite{Szczesniak2006B}. Slight difference in the shape of the discussed function occurs only in the vicinity of $\left<n\right>=1$, where characteristic maximum can be observed. Situation drastically changes for the high values of the EEPh potential ($v_{2}=5$ eV). The order parameter vanishes for $\left<n\right><1$. It also plots the curve whose shape is observed for cuprates \cite{Dagotto1994A} – even the characteristic plateau of the function $\Delta_{n=1}\left(\left<n\right>\right)$ is evident for $\left<n\right>\in\left<0.8,0.9\right>$ \cite{Liang2006A}. In the case of the wave function renormalization factor (Fig. \fig{f09} (B)), its values increase together with an increasing $v_{2}$. They are not, however, so heavily dependent on the doping as the order parameter. Let us note that the relatively weak dependence of $Z_{n=1}$ on $\left<n\right>$ is confirmed by the experimental values of the electron effective mass obtained for the superconductors ${\rm La_{2-x}Sr_{x}CuO_{4}}$ and ${\rm YBa_{2}Cu_{3}O_{y}}$ \cite{Padilla2005A}. The figure \fig{f09} (C) presents the plot of the energy shift function. Results prove that the strongest renormalization of the electron band energy exists in the area of low $\left<n\right>$ at the high values of $v_{2}$.
%
%

\begin{figure*}
\includegraphics[scale=0.60]{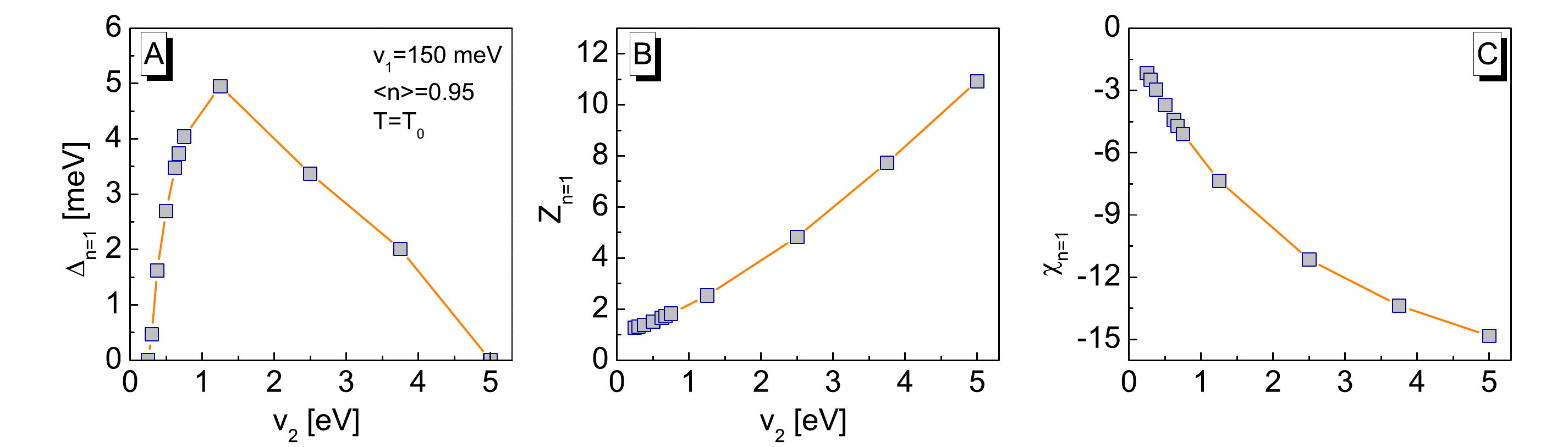}
\caption{
(A) The maximum value of order parameter, (B) the wave function renormalization factor, and (C) the energy shift function as a function of $v_{2}$.
}
\label{f10}
\end{figure*}

Interesting results concerning the superconducting state can be obtained for doping $<n>=0.95$, in the case of which, the highest value of the critical temperature was determined ($T_{C}=39.4$ K). By studying the dependence of the order parameter on the EEPh potential, we found that initially increasing $v_{2}$ causes the strong increase of $\Delta_{n=1}$. Nevertheless, above $1.25$ eV, the order parameter again starts to decrease and disappear for $v_{2}=5$ eV (Fig. \fig{f10} (A)). From the physical point of view, this means that the highest value of the critical temperature is observed for the relatively weak EEPh interaction. It turns out that the cause of the decrease of the critical temperature in the range of the high values of $v_{2}$ is connected with significant increase of the electron effective mass and the strong renormalization of the electron band energy (Fig. \fig{f10} (B) and (C)). 
%
%

\begin{figure*}
\includegraphics[scale=0.60]{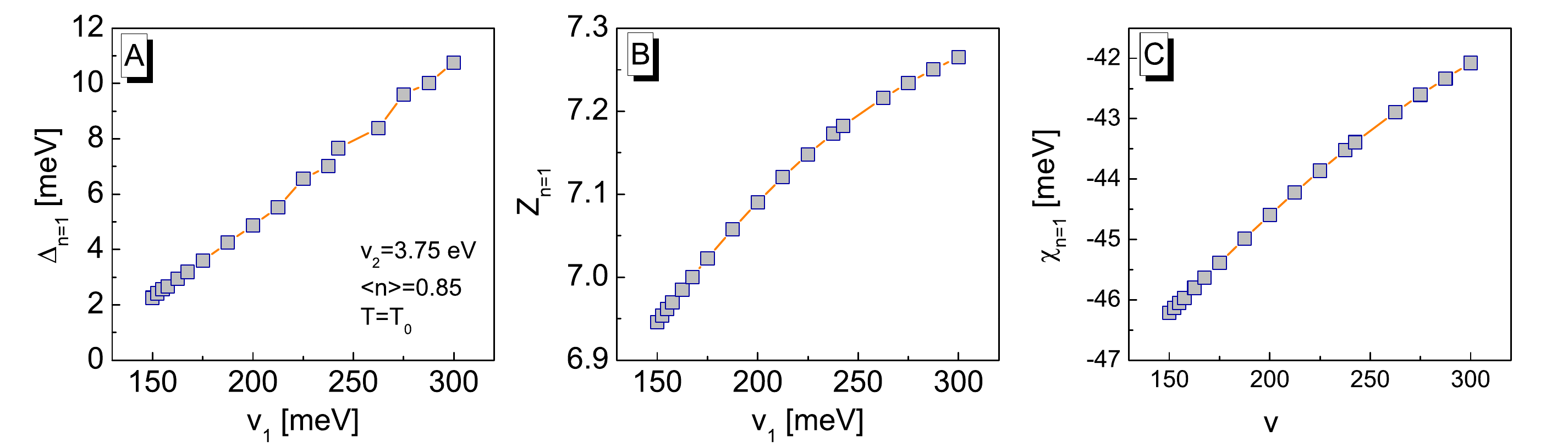}
\caption{
(A) The maximum value of order parameter, (B) the wave function renormalization factor, and (C) the band energy shift function as a function of $v_{1}$.
}
\label{f11}
\end{figure*}

The dependence of the order parameter on the potential $v_{1}$ looks completely different. For any value of doping, for which the superconducting state is observed, increase in $v_{1}$ always causes the increase in $\Delta_{n=1}$ (and therefore also increase in $T_{C}$). Sample results are presented in the figure \fig{f11} (A). Additionally, the figures \fig{f11} (B) and (C) show the plots of the wave function renormalization factor and the shift function. It can be seen that increasing potential $v_{1}$ does not result in the significant increase in $Z_{n=1}$, the decrease in absolute value of $\chi_{n=1}$ is also observed. 

Referring to the results of above discussion, it can be assumed that the superconducting state with the maximum critical temperature would rise in the case, in which EPh interaction will be strong enough, and EEPh interaction will be relatively weak – so that both the electron band mass and the electron band energy would not be renormalized in the significant manner. Taking into consideration the significant dependence of $v_{1}$ and $v_{2}$ on the distance between atoms building the crystal lattice, it is necessary to consider the possibility of carrying out the series of the high-pressure experiments – analogously as it was done for lithium, calcium, and hydrogenated compounds \cite{Deemyad2003A, Yabuuchi2006A, Sakata2011A, Strobel2010A, Eremets2008A}.
%
%

The Eliashberg equations on the imaginary axis enable to determine the exact value of the critical temperature. They also allow (with the good approximation) the calculation of order parameter, electron effective mass, and energy shift function. However, imaginary axis formalism is useless in the analysis of the superconducting state in the non-static area ($\omega\neq 0$). In order to determine the significance of dynamic effects, $\Delta_{n}$, $Z_{n}$, and $\chi_{n}$ should be analytically continued on real axis: $\Delta_{n}\rightarrow\Delta\left(\omega\right)$, $Z_{n}\rightarrow Z\left(\omega\right)$, and $\chi_{n}\rightarrow\chi\left(\omega\right)$. The easiest way to do it is using the formula \cite{Beach2000A}: 
\begin{equation}
\label{r1(5.2)}
X\left(\omega\right)=\frac{p_{1}+p_{2}\omega+...+p_{r}\omega^{r-1}}
{q_{1}+q_{2}\omega+...+q_{r}\omega^{r-1}+\omega^{r}},
\end{equation}
where: $X\in\left\{\Delta, Z, \chi \right\}$. Parameters $p_{j}$ and $q_{j}$ are determined in the numerical manner. In addition, we have adopted $r=6$.
%
\begin{figure*}
\includegraphics[scale=0.60]{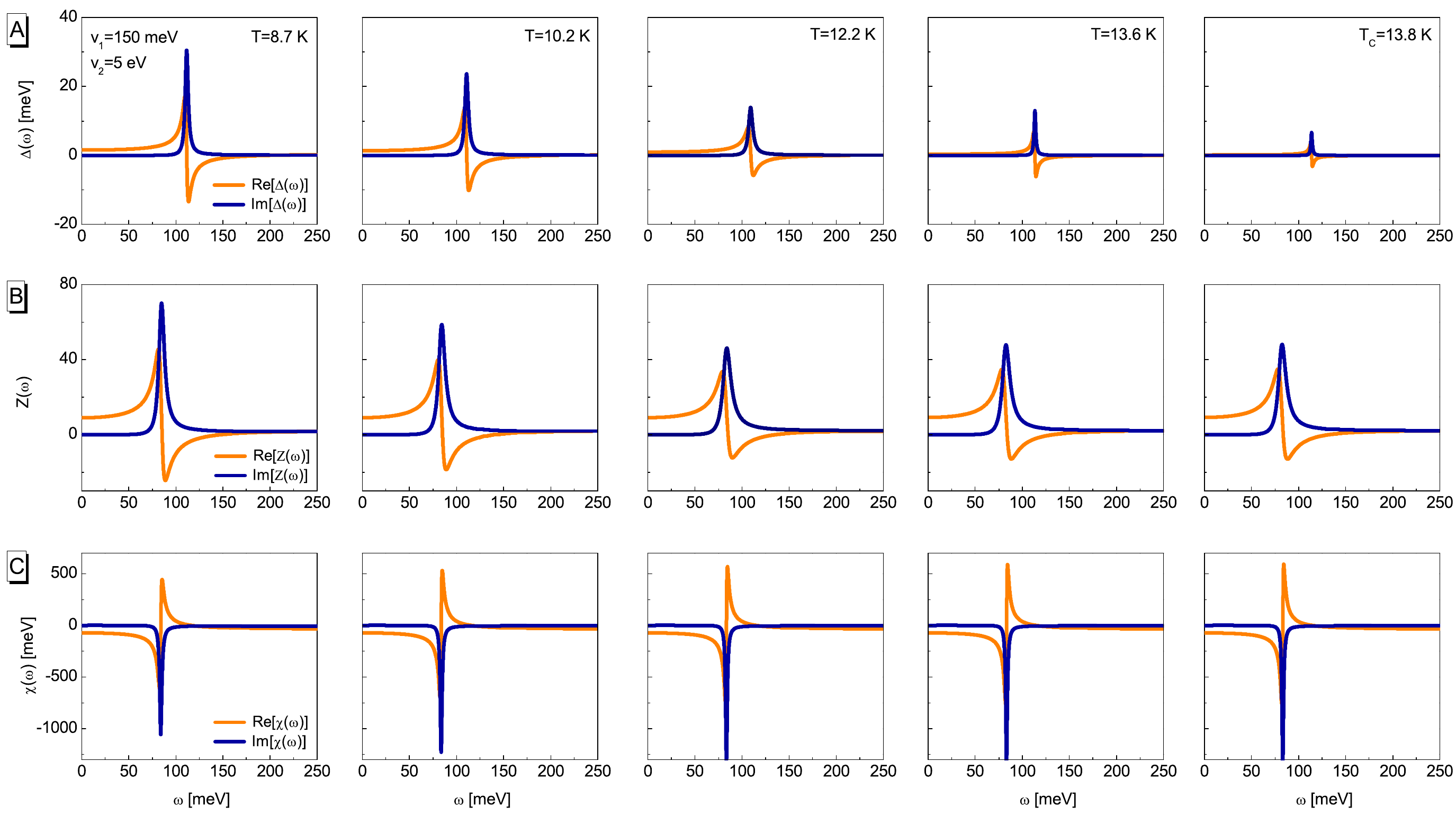}
\caption{
The form of the order parameter (A), the wave function renormalization factor (B), and the energy shift function (C) on the real axis. 
}
\label{f12}
\end{figure*}
%

The figures \fig{f12} (A)-(C) present exemplary results obtained for the case: $v_{1}=150$ meV, $v_{2}=5$ eV, and $\left<n\right>=0.8$, whereas the selected values of the temperature were taken into account. It can be seen that in the range of the low frequencies ($\omega\in\left<0,\sim 100\right>$ meV for $\Delta\left(\omega\right)$, and  $\omega\in\left<0,\sim 75\right>$ meV for $Z\left(\omega\right)$ and $\chi\left(\omega\right)$), the solutions of the Eliashberg equations take the real values. From the physical point of view, this means the lack of the damping effects \cite{Varelogiannis1997A}. In addition, in the considered frequency range, the real parts of solutions are poorly dependent on $\omega$, and it can be assumed that: ${\rm Re}\left[\Delta\left(\omega\right)\right]\simeq{\rm Re}\left[\Delta\left(0\right)\right]$, 
${\rm Re}\left[Z\left(\omega\right)\right]\simeq{\rm Re}\left[Z\left(0\right)\right]$, and 
${\rm Re}\left[\chi\left(\omega\right)\right]\simeq {\rm Re}\left[\chi\left(0\right)\right]$. The situation radically changes for $\omega\sim 110$ meV (in the case of $\Delta\left(\omega\right)$) and for $\omega\sim 80$ meV (in the case of $Z\left(\omega\right)$ and $\chi\left(\omega\right)$), where the appearance of the strong maximums and minimums can be observed. Let us note that above $150$ meV, the solutions of Eliashberg equations become extinguished. The increasing value of the temperature influences very differently on the evolution of $\Delta\left(\omega\right)$, $Z\left(\omega\right)$, and $\chi\left(\omega\right)$. It can be seen that the absolute values of the order parameter monotonically decrease with an increasing $T$. Interestingly, in the critical temperature even indicate non-zero values $\Delta\left(\omega\right)$ for $\omega\sim 110$ meV - these are the remains of the strong low-temperature maxima and minima. In the case of $Z\left(\omega\right)$ and $\chi\left(\omega\right)$, we practically do not observe the clear influence of the temperature on their low-frequency values, but their maxima and minima evolve differently - for the wave function renormalization factor they noticeably decrease and for the energy shift function they increase. It should be emphasized that the observed values of $Z\left(\omega\right)$ and $\chi\left(\omega\right)$ are significantly higher than the values of the functions in question, if we take into account only the electron-phonon interaction.
  
The physical values of the order parameter, the wave function renormalization factor, and the energy shift function should be calculated in the framework of the real axis Eliashberg formalism with the help of the equation \eq{r2(0)} and the following formulas: 
\begin{eqnarray}
\label{r2(5.2)}
Z\left(T\right)&=&{\rm Re}\left[Z\left(\omega=0,T\right)\right],\\ \nonumber 
\chi\left(T\right)&=&{\rm Re}\left[\chi\left(\omega=0,T\right)\right].
\end{eqnarray}

\begin{figure*}
\includegraphics[scale=0.55]{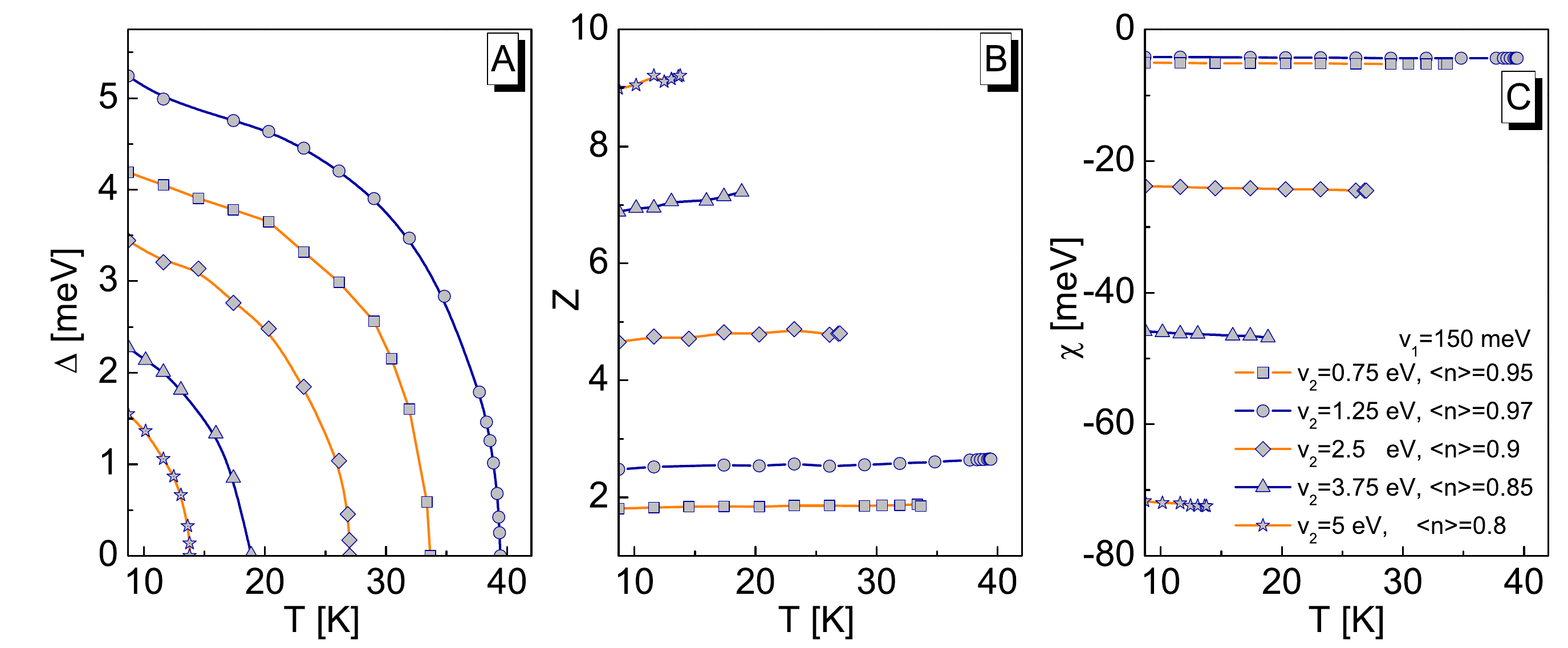}
\caption{
The physical value of the order parameter (A), the wave function renormalization factor (B), and the energy shift function (C). 
}
\label{f13}
\end{figure*}

The results obtained for the cases analogous as in the figure \fig{f08} are collected in the figure \fig{f13}. As expected, the differences between data obtained in the framework of the imaginary axis and the real axis formalism are very slight, which proves that the analytical continuation procedure was properly conducted.  
%
%

\begin{figure*}
\includegraphics[scale=0.55]{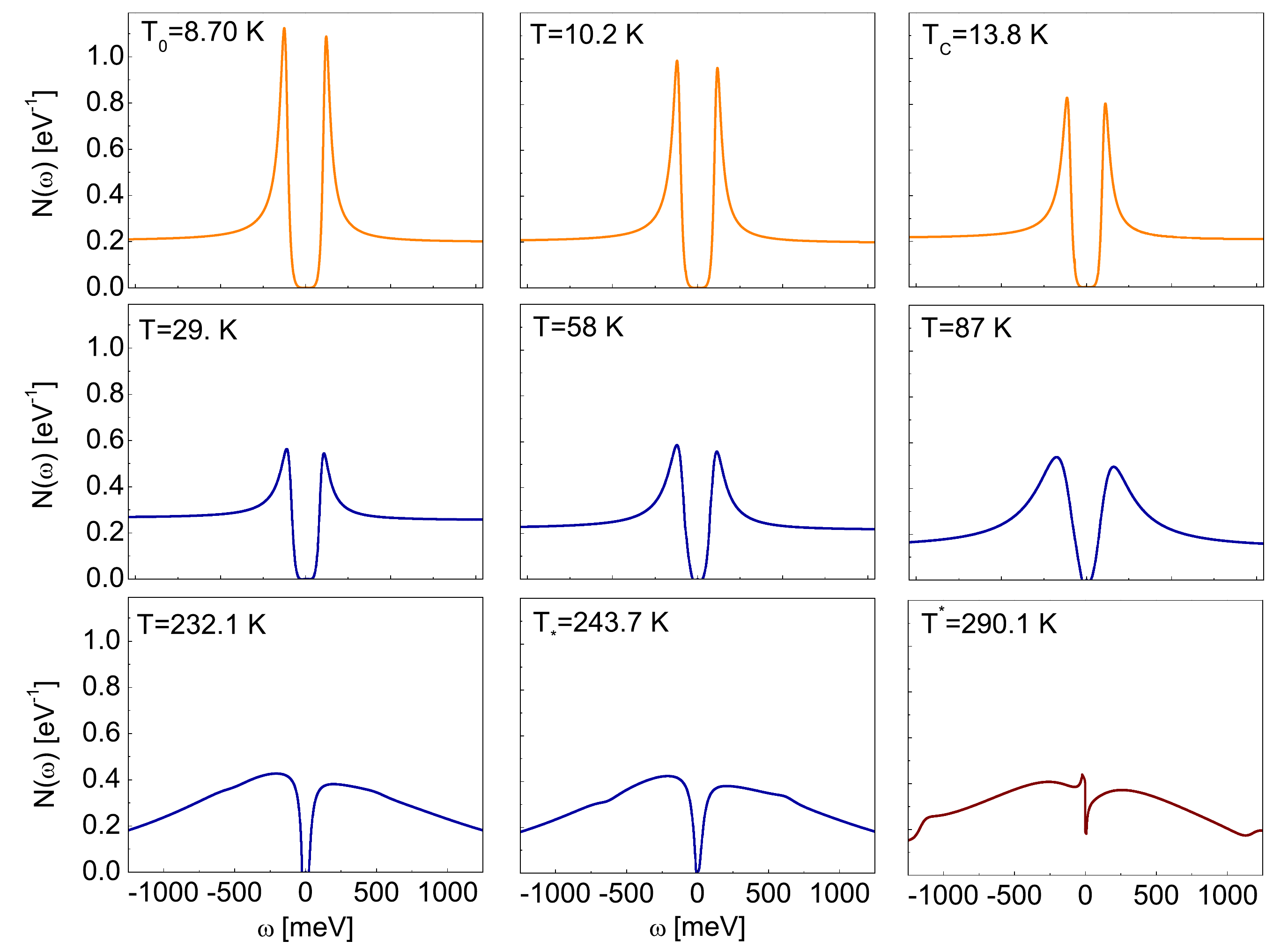}
\caption{
The temperature evolution of the electron density of states.
}
\label{f14}
\end{figure*}

As it was already mentioned, the real axis formalism allows to study the properties of the superconducting state outside the static area. It is possible then to give the answer to question: what can be the reason of the creation of the anomalous energy gap in the electron density of states, observed experimentally in cuprates \cite{Renner1998A, Renner1998B, Fischer2007A}. The figure \fig{f14} presents the form of $N\left(\omega\right)$ determined for the case: $v_{1}=150$ meV, $v_{2}=5$ eV, and $\left<n\right>=0.8$. It was found that in the range of the temperature from $T_{0}$ to $T^{\star}=290.1$ K, the energy gap induces in the electron density of states at the Fermi level. In particular, for temperatures from $T_{0}$ to $T_{C}$, the half-width of the energy gap changes slightly, while the clear decrease in the maximum value of $N\left(\omega\right)$ may be noticed. Interestingly, in contrast to the shape of the electron density of states, associated with classical electron-phonon interaction, the curve $N\left(\omega\right)$ is clearly non-symmetrical with respect to the Fermi level. The observed asymmetry results from the strong electron band energy renormalization, which is caused by the high values of $\chi\left(\omega\right)$. It should be noted that theoretically determined asymmetry of the electron density of states is very clearly visible in the shape of the experimental curves \cite{Renner1998A}. The energy gap wears off slowly above the critical temperature. The first non-zero value of the function $N\left(\omega\right)$ on the Fermi level can be observed only at the temperature $T_{\star}$ equal to $243.7$ K. For $T\geq T_{\star}$, in the course of the electron density of states still visible is the remnant of the energy gap, which gradually disappears with the increasing temperature. Finally, the pseudogap disappears at $T^{\star}$. The presented model predicts that gap in the electron density of states, in the full range of the temperature (even below $T_{C}$), is induced by $Z\left(\omega\right)$ and $\chi\left(\omega\right)$. The overt contribution of $N\left(\omega\right)$ lodged by the superconducting state is negligibly small. This can be easily checked determining the shape of the electron density of states on the basis of the formula \eq{r1(0)}, in which it was adopted $\varphi\left(\omega\right)=0$.
%
\vspace{0.5cm}
\section{Summary}
In the presented work we have partially reproduced the properties of the superconducting and the anomalous normal state in cuprates. Considerations have been based on the Hamiltonian modeling the electron-phonon and the electron-electron-phonon interaction. The thermodynamic equations have been derived in the framework of the Eliashberg formalism. We have shown that the experimental dependence of the order parameter on the doping can be qualitatively reproduced for the sufficiently large value of the EEPh potential. Then, it was proven that the Eliashberg equations correctly characterize the type of the normal-superconducting state phase transition. In our opinion, the most interesting result was obtained during explaining the origin of the energy gap in the electron density of states. We have shown that the energy gap is induced by the diagonal elements of the matrix self-energy ($Z$ and $\chi$), which are associated with the anomalous normal state above $T_{C}$. The explicit contribution to the course of the electron density of states derived from superconducting state proves to be negligibly small.
  
%
%
\bibliography{Bibliography}

\begin{thebibliography}{63}
\expandafter\ifx\csname natexlab\endcsname\relax\def\natexlab#1{#1}\fi
\expandafter\ifx\csname bibnamefont\endcsname\relax
  \def\bibnamefont#1{#1}\fi
\expandafter\ifx\csname bibfnamefont\endcsname\relax
  \def\bibfnamefont#1{#1}\fi
\expandafter\ifx\csname citenamefont\endcsname\relax
  \def\citenamefont#1{#1}\fi
\expandafter\ifx\csname url\endcsname\relax
  \def\url#1{\texttt{#1}}\fi
\expandafter\ifx\csname urlprefix\endcsname\relax\def\urlprefix{URL }\fi
\providecommand{\bibinfo}[2]{#2}
\providecommand{\eprint}[2][]{\url{#2}}

\bibitem[{\citenamefont{Fr{\"o}hlich}(1950)}]{Frohlich1950A}
\bibinfo{author}{\bibfnamefont{H.}~\bibnamefont{Fr{\"o}hlich}},
  \bibinfo{journal}{Physical Review} \textbf{\bibinfo{volume}{79}},
  \bibinfo{pages}{845} (\bibinfo{year}{1950}).

\bibitem[{\citenamefont{Fr{\"o}hlich}(1952)}]{Frohlich1952A}
\bibinfo{author}{\bibfnamefont{H.}~\bibnamefont{Fr{\"o}hlich}},
  \bibinfo{journal}{Proceedings of the Royal Society of London A}
  \textbf{\bibinfo{volume}{215}}, \bibinfo{pages}{291} (\bibinfo{year}{1952}).

\bibitem[{\citenamefont{Fr{\"o}hlich}(1954)}]{Frohlich1954A}
\bibinfo{author}{\bibfnamefont{H.}~\bibnamefont{Fr{\"o}hlich}},
  \bibinfo{journal}{Proceedings of the Royal Society of London A}
  \textbf{\bibinfo{volume}{223}}, \bibinfo{pages}{296} (\bibinfo{year}{1954}).

\bibitem[{\citenamefont{Bardeen
  et~al.}(1957{\natexlab{a}})\citenamefont{Bardeen, Cooper, and
  Schrieffer}}]{Bardeen1957A}
\bibinfo{author}{\bibfnamefont{J.}~\bibnamefont{Bardeen}},
  \bibinfo{author}{\bibfnamefont{L.~N.} \bibnamefont{Cooper}},
  \bibnamefont{and} \bibinfo{author}{\bibfnamefont{J.~R.}
  \bibnamefont{Schrieffer}}, \bibinfo{journal}{Physical Review}
  \textbf{\bibinfo{volume}{106}}, \bibinfo{pages}{162}
  (\bibinfo{year}{1957}{\natexlab{a}}).

\bibitem[{\citenamefont{Bardeen
  et~al.}(1957{\natexlab{b}})\citenamefont{Bardeen, Cooper, and
  Schrieffer}}]{Bardeen1957B}
\bibinfo{author}{\bibfnamefont{J.}~\bibnamefont{Bardeen}},
  \bibinfo{author}{\bibfnamefont{L.~N.} \bibnamefont{Cooper}},
  \bibnamefont{and} \bibinfo{author}{\bibfnamefont{J.~R.}
  \bibnamefont{Schrieffer}}, \bibinfo{journal}{Physical Review}
  \textbf{\bibinfo{volume}{108}}, \bibinfo{pages}{1175}
  (\bibinfo{year}{1957}{\natexlab{b}}).

\bibitem[{\citenamefont{Fetter and Walecka}(1971)}]{Fetter1971A}
\bibinfo{author}{\bibfnamefont{A.~L.} \bibnamefont{Fetter}} \bibnamefont{and}
  \bibinfo{author}{\bibfnamefont{J.~D.} \bibnamefont{Walecka}},
  \emph{\bibinfo{title}{Quantum Theory of Many-Particle Systems}}
  (\bibinfo{publisher}{McGraw-Hill Book Company}, \bibinfo{year}{1971}).

\bibitem[{\citenamefont{Carbotte}(1990)}]{Carbotte1990A}
\bibinfo{author}{\bibfnamefont{J.~P.} \bibnamefont{Carbotte}},
  \bibinfo{journal}{Reviews of Modern Physics} \textbf{\bibinfo{volume}{62}},
  \bibinfo{pages}{1027} (\bibinfo{year}{1990}).

\bibitem[{\citenamefont{Carbotte and Marsiglio}(2003)}]{Carbotte2003A}
\bibinfo{author}{\bibfnamefont{J.~P.} \bibnamefont{Carbotte}} \bibnamefont{and}
  \bibinfo{author}{\bibfnamefont{F.}~\bibnamefont{Marsiglio}}, in
  \emph{\bibinfo{booktitle}{The Physics of Superconductors edited by K. H.
  Bennemann and J. B. Ketterson}} (\bibinfo{publisher}{Springer Berlin
  Heidelberg}, \bibinfo{year}{2003}).

\bibitem[{\citenamefont{Eliashberg}(1960)}]{Eliashberg1960A}
\bibinfo{author}{\bibfnamefont{G.~M.} \bibnamefont{Eliashberg}},
  \bibinfo{journal}{Soviet Physics JETP} \textbf{\bibinfo{volume}{11}},
  \bibinfo{pages}{696} (\bibinfo{year}{1960}).

\bibitem[{\citenamefont{Gasser et~al.}(1999)\citenamefont{Gasser, Heiner, and
  Elk}}]{Gasser1999A}
\bibinfo{author}{\bibfnamefont{W.}~\bibnamefont{Gasser}},
  \bibinfo{author}{\bibfnamefont{E.}~\bibnamefont{Heiner}}, \bibnamefont{and}
  \bibinfo{author}{\bibfnamefont{K.}~\bibnamefont{Elk}},
  \emph{\bibinfo{title}{Greensche Funktionen in Festk{\"o}rper- und
  Vielteilchenphysik}} (\bibinfo{publisher}{VILEY-VCH Verlag GmbH},
  \bibinfo{address}{Weinheim}, \bibinfo{year}{1999}).

\bibitem[{\citenamefont{Szcz{\c{e}}{\'s}niak}(2007{\natexlab{a}})}]{Szczesniak2007A}
\bibinfo{author}{\bibfnamefont{R.}~\bibnamefont{Szcz{\c{e}}{\'s}niak}},
  \bibinfo{journal}{Physica Status Solidi B} \textbf{\bibinfo{volume}{244}},
  \bibinfo{pages}{2538} (\bibinfo{year}{2007}{\natexlab{a}}).

\bibitem[{\citenamefont{Szcz{\c{e}}{\'s}niak}(2007{\natexlab{b}})}]{Szczesniak2007B}
\bibinfo{author}{\bibfnamefont{R.}~\bibnamefont{Szcz{\c{e}}{\'s}niak}},
  \bibinfo{journal}{Solid State Communications} \textbf{\bibinfo{volume}{144}},
  \bibinfo{pages}{27} (\bibinfo{year}{2007}{\natexlab{b}}).

\bibitem[{\citenamefont{Szcz{\c{e}}{\'s}niak}(2008)}]{Szczesniak2008A}
\bibinfo{author}{\bibfnamefont{R.}~\bibnamefont{Szcz{\c{e}}{\'s}niak}},
  \bibinfo{journal}{Solid State Communications} \textbf{\bibinfo{volume}{145}},
  \bibinfo{pages}{137} (\bibinfo{year}{2008}).

\bibitem[{\citenamefont{Szcz{\c{e}}{\'s}niak
  et~al.}(2010)\citenamefont{Szcz{\c{e}}{\'s}niak, Jarosik, and
  Szcz{\c{e}}{\'s}niak}}]{Szczesniak2010A}
\bibinfo{author}{\bibfnamefont{R.}~\bibnamefont{Szcz{\c{e}}{\'s}niak}},
  \bibinfo{author}{\bibfnamefont{M.~W.} \bibnamefont{Jarosik}},
  \bibnamefont{and}
  \bibinfo{author}{\bibfnamefont{D.}~\bibnamefont{Szcz{\c{e}}{\'s}niak}},
  \bibinfo{journal}{Physica B} \textbf{\bibinfo{volume}{405}},
  \bibinfo{pages}{4897} (\bibinfo{year}{2010}).

\bibitem[{\citenamefont{Bednorz and M{\" u}ller}(1986)}]{Bednorz1986A}
\bibinfo{author}{\bibfnamefont{J.~G.} \bibnamefont{Bednorz}} \bibnamefont{and}
  \bibinfo{author}{\bibfnamefont{K.~A.} \bibnamefont{M{\" u}ller}},
  \bibinfo{journal}{Zeitschrift f{\" u}r Physik B Condensed Matter}
  \textbf{\bibinfo{volume}{64}}, \bibinfo{pages}{189} (\bibinfo{year}{1986}).

\bibitem[{\citenamefont{Bednorz and M{\"u}ller}(1988)}]{Bednorz1988A}
\bibinfo{author}{\bibfnamefont{J.~G.} \bibnamefont{Bednorz}} \bibnamefont{and}
  \bibinfo{author}{\bibfnamefont{K.~A.} \bibnamefont{M{\"u}ller}},
  \bibinfo{journal}{Reviews of Modern Physics} \textbf{\bibinfo{volume}{60}},
  \bibinfo{pages}{585} (\bibinfo{year}{1988}).

\bibitem[{\citenamefont{Dagotto}(1994)}]{Dagotto1994A}
\bibinfo{author}{\bibfnamefont{E.}~\bibnamefont{Dagotto}},
  \bibinfo{journal}{Reviews of Modern Physics} \textbf{\bibinfo{volume}{66}},
  \bibinfo{pages}{763} (\bibinfo{year}{1994}).

\bibitem[{\citenamefont{King et~al.}(1995)\citenamefont{King, Dessau, Loeser,
  Shen, and Wells}}]{King1995A}
\bibinfo{author}{\bibfnamefont{D.~M.} \bibnamefont{King}},
  \bibinfo{author}{\bibfnamefont{D.~S.} \bibnamefont{Dessau}},
  \bibinfo{author}{\bibfnamefont{A.~G.} \bibnamefont{Loeser}},
  \bibinfo{author}{\bibfnamefont{Z.~X.} \bibnamefont{Shen}}, \bibnamefont{and}
  \bibinfo{author}{\bibfnamefont{B.~O.} \bibnamefont{Wells}},
  \bibinfo{journal}{Journal of Physics and Chemistry of Solids}
  \textbf{\bibinfo{volume}{56}}, \bibinfo{pages}{1865} (\bibinfo{year}{1995}).

\bibitem[{\citenamefont{Renner et~al.}(1998{\natexlab{a}})\citenamefont{Renner,
  Revaz, Genoud, Kadowaki, and Fischer}}]{Renner1998A}
\bibinfo{author}{\bibfnamefont{C.}~\bibnamefont{Renner}},
  \bibinfo{author}{\bibfnamefont{B.}~\bibnamefont{Revaz}},
  \bibinfo{author}{\bibfnamefont{J.~Y.} \bibnamefont{Genoud}},
  \bibinfo{author}{\bibfnamefont{K.}~\bibnamefont{Kadowaki}}, \bibnamefont{and}
  \bibinfo{author}{\bibfnamefont{O.}~\bibnamefont{Fischer}},
  \bibinfo{journal}{Physical Review Letters} \textbf{\bibinfo{volume}{80}},
  \bibinfo{pages}{149} (\bibinfo{year}{1998}{\natexlab{a}}).

\bibitem[{\citenamefont{Renner et~al.}(1998{\natexlab{b}})\citenamefont{Renner,
  Revaz, Kadowaki, Maggio-Aprile, and Fischer}}]{Renner1998B}
\bibinfo{author}{\bibfnamefont{C.}~\bibnamefont{Renner}},
  \bibinfo{author}{\bibfnamefont{B.}~\bibnamefont{Revaz}},
  \bibinfo{author}{\bibfnamefont{K.}~\bibnamefont{Kadowaki}},
  \bibinfo{author}{\bibfnamefont{I.}~\bibnamefont{Maggio-Aprile}},
  \bibnamefont{and} \bibinfo{author}{\bibfnamefont{O.}~\bibnamefont{Fischer}},
  \bibinfo{journal}{Physical Review Letters} \textbf{\bibinfo{volume}{80}},
  \bibinfo{pages}{3606} (\bibinfo{year}{1998}{\natexlab{b}}).

\bibitem[{\citenamefont{Fischer et~al.}(2007)\citenamefont{Fischer, Kugler,
  Maggio-Aprile, and Berthod}}]{Fischer2007A}
\bibinfo{author}{\bibfnamefont{O.}~\bibnamefont{Fischer}},
  \bibinfo{author}{\bibfnamefont{M.}~\bibnamefont{Kugler}},
  \bibinfo{author}{\bibfnamefont{I.}~\bibnamefont{Maggio-Aprile}},
  \bibnamefont{and} \bibinfo{author}{\bibfnamefont{C.}~\bibnamefont{Berthod}},
  \bibinfo{journal}{Reviews of Modern Physics} \textbf{\bibinfo{volume}{79}},
  \bibinfo{pages}{353} (\bibinfo{year}{2007}).

\bibitem[{\citenamefont{Marhall et~al.}(1996)\citenamefont{Marhall, Dessau,
  Loeser, Park, Matsuura, Eckstein, Bozovic, Fournier, Kapitulnik, Spicer
  et~al.}}]{Marshall1996A}
\bibinfo{author}{\bibfnamefont{D.~S.} \bibnamefont{Marhall}},
  \bibinfo{author}{\bibfnamefont{D.~S.} \bibnamefont{Dessau}},
  \bibinfo{author}{\bibfnamefont{A.~G.} \bibnamefont{Loeser}},
  \bibinfo{author}{\bibfnamefont{C.~H.} \bibnamefont{Park}},
  \bibinfo{author}{\bibfnamefont{A.~Y.} \bibnamefont{Matsuura}},
  \bibinfo{author}{\bibfnamefont{J.~N.} \bibnamefont{Eckstein}},
  \bibinfo{author}{\bibfnamefont{I.}~\bibnamefont{Bozovic}},
  \bibinfo{author}{\bibfnamefont{P.}~\bibnamefont{Fournier}},
  \bibinfo{author}{\bibfnamefont{A.}~\bibnamefont{Kapitulnik}},
  \bibinfo{author}{\bibfnamefont{W.~E.} \bibnamefont{Spicer}},
  \bibnamefont{et~al.}, \bibinfo{journal}{Physical Review Letters}
  \textbf{\bibinfo{volume}{76}}, \bibinfo{pages}{4841} (\bibinfo{year}{1996}).

\bibitem[{\citenamefont{Loeser et~al.}(1996)\citenamefont{Loeser, Shen, Dessau,
  Marhall, Park, Fournier, and Kapitulnik}}]{Loeser1996A}
\bibinfo{author}{\bibfnamefont{A.~G.} \bibnamefont{Loeser}},
  \bibinfo{author}{\bibfnamefont{Z.~X.} \bibnamefont{Shen}},
  \bibinfo{author}{\bibfnamefont{D.~S.} \bibnamefont{Dessau}},
  \bibinfo{author}{\bibfnamefont{D.~S.} \bibnamefont{Marhall}},
  \bibinfo{author}{\bibfnamefont{C.~H.} \bibnamefont{Park}},
  \bibinfo{author}{\bibfnamefont{P.}~\bibnamefont{Fournier}}, \bibnamefont{and}
  \bibinfo{author}{\bibfnamefont{A.}~\bibnamefont{Kapitulnik}},
  \bibinfo{journal}{Science} \textbf{\bibinfo{volume}{273}},
  \bibinfo{pages}{325} (\bibinfo{year}{1996}).

\bibitem[{\citenamefont{Ding et~al.}(1996)\citenamefont{Ding, Norman,
  Campuzano, Randeria, Bellman, Yokoya, Takahashi, Mochiku, and
  Kadowski}}]{Ding1996A}
\bibinfo{author}{\bibfnamefont{H.~M.} \bibnamefont{Ding}},
  \bibinfo{author}{\bibfnamefont{M.~R.} \bibnamefont{Norman}},
  \bibinfo{author}{\bibfnamefont{J.~C.} \bibnamefont{Campuzano}},
  \bibinfo{author}{\bibfnamefont{M.}~\bibnamefont{Randeria}},
  \bibinfo{author}{\bibfnamefont{A.~F.} \bibnamefont{Bellman}},
  \bibinfo{author}{\bibfnamefont{T.}~\bibnamefont{Yokoya}},
  \bibinfo{author}{\bibfnamefont{T.}~\bibnamefont{Takahashi}},
  \bibinfo{author}{\bibfnamefont{T.}~\bibnamefont{Mochiku}}, \bibnamefont{and}
  \bibinfo{author}{\bibfnamefont{K.}~\bibnamefont{Kadowski}},
  \bibinfo{journal}{Physical Review B} \textbf{\bibinfo{volume}{54}},
  \bibinfo{pages}{9678} (\bibinfo{year}{1996}).

\bibitem[{\citenamefont{Damascelli et~al.}(2003)\citenamefont{Damascelli,
  Hussain, and Shen}}]{Damascelli2003A}
\bibinfo{author}{\bibfnamefont{A.}~\bibnamefont{Damascelli}},
  \bibinfo{author}{\bibfnamefont{Z.}~\bibnamefont{Hussain}}, \bibnamefont{and}
  \bibinfo{author}{\bibfnamefont{Z.~X.} \bibnamefont{Shen}},
  \bibinfo{journal}{Reviews of Modern Physics} \textbf{\bibinfo{volume}{75}},
  \bibinfo{pages}{473} (\bibinfo{year}{2003}).

\bibitem[{\citenamefont{Emery}(1987)}]{Emery1987A}
\bibinfo{author}{\bibfnamefont{V.~J.} \bibnamefont{Emery}},
  \bibinfo{journal}{Physical Review Letters} \textbf{\bibinfo{volume}{58}},
  \bibinfo{pages}{2794} (\bibinfo{year}{1987}).

\bibitem[{\citenamefont{Littlewood et~al.}(1989)\citenamefont{Littlewood,
  Varma, and Abrahams}}]{Littlewood1989A}
\bibinfo{author}{\bibfnamefont{P.~B.} \bibnamefont{Littlewood}},
  \bibinfo{author}{\bibfnamefont{C.~M.} \bibnamefont{Varma}}, \bibnamefont{and}
  \bibinfo{author}{\bibfnamefont{E.}~\bibnamefont{Abrahams}},
  \bibinfo{journal}{Physical Review Letters} \textbf{\bibinfo{volume}{60}},
  \bibinfo{pages}{2602} (\bibinfo{year}{1989}).

\bibitem[{\citenamefont{Hybertsen et~al.}(1990)\citenamefont{Hybertsen,
  Stechel, Schluter, and Jennison}}]{Hybertsen1990A}
\bibinfo{author}{\bibfnamefont{M.~S.} \bibnamefont{Hybertsen}},
  \bibinfo{author}{\bibfnamefont{E.~B.} \bibnamefont{Stechel}},
  \bibinfo{author}{\bibfnamefont{M.}~\bibnamefont{Schluter}}, \bibnamefont{and}
  \bibinfo{author}{\bibfnamefont{D.~R.} \bibnamefont{Jennison}},
  \bibinfo{journal}{Physical Review B} \textbf{\bibinfo{volume}{41}},
  \bibinfo{pages}{11068} (\bibinfo{year}{1990}).

\bibitem[{\citenamefont{Cuk et~al.}(2005)\citenamefont{Cuk, Lu, Zhou, Shen,
  Deveraux, and Nagaosa}}]{Cuk2005A}
\bibinfo{author}{\bibfnamefont{T.}~\bibnamefont{Cuk}},
  \bibinfo{author}{\bibfnamefont{D.~H.} \bibnamefont{Lu}},
  \bibinfo{author}{\bibfnamefont{X.~J.} \bibnamefont{Zhou}},
  \bibinfo{author}{\bibfnamefont{Z.~X.} \bibnamefont{Shen}},
  \bibinfo{author}{\bibfnamefont{T.~P.} \bibnamefont{Deveraux}},
  \bibnamefont{and} \bibinfo{author}{\bibfnamefont{N.}~\bibnamefont{Nagaosa}},
  \bibinfo{journal}{Physica Status Solidi B} \textbf{\bibinfo{volume}{242}},
  \bibinfo{pages}{11} (\bibinfo{year}{2005}).

\bibitem[{\citenamefont{Gweon et~al.}(2004)\citenamefont{Gweon, Sasagawa, Zhou,
  Graf, Takagi, Lee, and Lanzara}}]{Gweon2004A}
\bibinfo{author}{\bibfnamefont{G.~H.} \bibnamefont{Gweon}},
  \bibinfo{author}{\bibfnamefont{T.}~\bibnamefont{Sasagawa}},
  \bibinfo{author}{\bibfnamefont{S.~Y.} \bibnamefont{Zhou}},
  \bibinfo{author}{\bibfnamefont{J.}~\bibnamefont{Graf}},
  \bibinfo{author}{\bibfnamefont{H.}~\bibnamefont{Takagi}},
  \bibinfo{author}{\bibfnamefont{D.~H.} \bibnamefont{Lee}}, \bibnamefont{and}
  \bibinfo{author}{\bibfnamefont{A.}~\bibnamefont{Lanzara}},
  \bibinfo{journal}{Nature} \textbf{\bibinfo{volume}{430}},
  \bibinfo{pages}{187} (\bibinfo{year}{2004}).

\bibitem[{\citenamefont{Douglas et~al.}(2007)\citenamefont{Douglas, Iwasawa,
  Sun, Fedorov, Ishikado, Saitoh, Eisaki, Bando, Iwase, Ino
  et~al.}}]{Douglas2007A}
\bibinfo{author}{\bibfnamefont{J.~F.} \bibnamefont{Douglas}},
  \bibinfo{author}{\bibfnamefont{H.}~\bibnamefont{Iwasawa}},
  \bibinfo{author}{\bibfnamefont{Z.}~\bibnamefont{Sun}},
  \bibinfo{author}{\bibfnamefont{A.~V.} \bibnamefont{Fedorov}},
  \bibinfo{author}{\bibfnamefont{M.}~\bibnamefont{Ishikado}},
  \bibinfo{author}{\bibfnamefont{T.}~\bibnamefont{Saitoh}},
  \bibinfo{author}{\bibfnamefont{H.}~\bibnamefont{Eisaki}},
  \bibinfo{author}{\bibfnamefont{H.}~\bibnamefont{Bando}},
  \bibinfo{author}{\bibfnamefont{T.}~\bibnamefont{Iwase}},
  \bibinfo{author}{\bibfnamefont{A.}~\bibnamefont{Ino}}, \bibnamefont{et~al.},
  \bibinfo{journal}{Nature} \textbf{\bibinfo{volume}{446}}, \bibinfo{pages}{E5}
  (\bibinfo{year}{2007}).

\bibitem[{\citenamefont{Franck}(1994)}]{Franck1994A}
\bibinfo{author}{\bibfnamefont{J.~P.} \bibnamefont{Franck}},
  \emph{\bibinfo{title}{Physical Properties of High Temperature
  Superconductors}} (\bibinfo{publisher}{World Scientific},
  \bibinfo{address}{Singapore}, \bibinfo{year}{1994}).

\bibitem[{\citenamefont{Lee et~al.}(2006)\citenamefont{Lee, Fujita, McElroy,
  Slezak, Wang, Aiura, Bando, Ishikado, Masui, Zhu et~al.}}]{Lee2006A}
\bibinfo{author}{\bibfnamefont{J.}~\bibnamefont{Lee}},
  \bibinfo{author}{\bibfnamefont{K.}~\bibnamefont{Fujita}},
  \bibinfo{author}{\bibfnamefont{K.}~\bibnamefont{McElroy}},
  \bibinfo{author}{\bibfnamefont{J.~A.} \bibnamefont{Slezak}},
  \bibinfo{author}{\bibfnamefont{M.}~\bibnamefont{Wang}},
  \bibinfo{author}{\bibfnamefont{T.}~\bibnamefont{Aiura}},
  \bibinfo{author}{\bibfnamefont{H.}~\bibnamefont{Bando}},
  \bibinfo{author}{\bibfnamefont{M.}~\bibnamefont{Ishikado}},
  \bibinfo{author}{\bibfnamefont{T.}~\bibnamefont{Masui}},
  \bibinfo{author}{\bibfnamefont{J.~X.} \bibnamefont{Zhu}},
  \bibnamefont{et~al.}, \bibinfo{journal}{Nature}
  \textbf{\bibinfo{volume}{442}}, \bibinfo{pages}{546} (\bibinfo{year}{2006}).

\bibitem[{\citenamefont{Hofer et~al.}(2000)\citenamefont{Hofer, Conder,
  Sasagawa, Zhao, Willemin, Keller, and Kishio}}]{Hofer2000A}
\bibinfo{author}{\bibfnamefont{J.}~\bibnamefont{Hofer}},
  \bibinfo{author}{\bibfnamefont{K.}~\bibnamefont{Conder}},
  \bibinfo{author}{\bibfnamefont{T.}~\bibnamefont{Sasagawa}},
  \bibinfo{author}{\bibfnamefont{G.}~\bibnamefont{Zhao}},
  \bibinfo{author}{\bibfnamefont{M.}~\bibnamefont{Willemin}},
  \bibinfo{author}{\bibfnamefont{H.}~\bibnamefont{Keller}}, \bibnamefont{and}
  \bibinfo{author}{\bibfnamefont{K.}~\bibnamefont{Kishio}},
  \bibinfo{journal}{Physical Review Letters} \textbf{\bibinfo{volume}{84}},
  \bibinfo{pages}{4192} (\bibinfo{year}{2000}).

\bibitem[{\citenamefont{Schneider}(2005)}]{Schneider2005A}
\bibinfo{author}{\bibfnamefont{T.}~\bibnamefont{Schneider}},
  \bibinfo{journal}{Physica Status Solidi B} \textbf{\bibinfo{volume}{242}},
  \bibinfo{pages}{58} (\bibinfo{year}{2005}).

\bibitem[{\citenamefont{Kim and Tesanovic}(1993)}]{Kim1993A}
\bibinfo{author}{\bibfnamefont{J.~H.} \bibnamefont{Kim}} \bibnamefont{and}
  \bibinfo{author}{\bibfnamefont{Z.}~\bibnamefont{Tesanovic}},
  \bibinfo{journal}{Physical Review Letters} \textbf{\bibinfo{volume}{71}},
  \bibinfo{pages}{4218} (\bibinfo{year}{1993}).

\bibitem[{\citenamefont{Schneider et~al.}(2004)\citenamefont{Schneider,
  Hammerl, Logvenov, Kopp, Kirtley, Hirschfeld, and Mannhart}}]{Schneider2004A}
\bibinfo{author}{\bibfnamefont{C.~W.} \bibnamefont{Schneider}},
  \bibinfo{author}{\bibfnamefont{G.}~\bibnamefont{Hammerl}},
  \bibinfo{author}{\bibfnamefont{G.}~\bibnamefont{Logvenov}},
  \bibinfo{author}{\bibfnamefont{T.}~\bibnamefont{Kopp}},
  \bibinfo{author}{\bibfnamefont{J.~R.} \bibnamefont{Kirtley}},
  \bibinfo{author}{\bibfnamefont{P.~J.} \bibnamefont{Hirschfeld}},
  \bibnamefont{and} \bibinfo{author}{\bibfnamefont{J.}~\bibnamefont{Mannhart}},
  \bibinfo{journal}{Europhysics Letters} \textbf{\bibinfo{volume}{68}},
  \bibinfo{pages}{86} (\bibinfo{year}{2004}).

\bibitem[{\citenamefont{Szcz{\c{e}}{\'s}niak}(2012{\natexlab{a}})}]{Szczesniak2012G}
\bibinfo{author}{\bibfnamefont{R.}~\bibnamefont{Szcz{\c{e}}{\'s}niak}},
  \bibinfo{journal}{PloS ONE} \textbf{\bibinfo{volume}{7}},
  \bibinfo{pages}{e31873} (\bibinfo{year}{2012}{\natexlab{a}}).

\bibitem[{\citenamefont{Szcz{\c{e}}{\'s}niak}(2011)}]{Szczesniak2011D}
\bibinfo{author}{\bibfnamefont{R.}~\bibnamefont{Szcz{\c{e}}{\'s}niak}},
  \bibinfo{journal}{arXiv:1105.5525}  (\bibinfo{year}{2011}).

\bibitem[{\citenamefont{Rickayzen}(1965)}]{Rickayzen1964A}
\bibinfo{author}{\bibfnamefont{G.}~\bibnamefont{Rickayzen}},
  \emph{\bibinfo{title}{Theory of Superconductivity}}
  (\bibinfo{publisher}{Interscience}, \bibinfo{year}{1965}).

\bibitem[{\citenamefont{Ma{\'c}kowiak and Tarasewicz}(1998)}]{Mackowiak1998A}
\bibinfo{author}{\bibfnamefont{J.}~\bibnamefont{Ma{\'c}kowiak}}
  \bibnamefont{and}
  \bibinfo{author}{\bibfnamefont{P.}~\bibnamefont{Tarasewicz}},
  \bibinfo{journal}{Acta Physica Polonica A} \textbf{\bibinfo{volume}{93}},
  \bibinfo{pages}{659} (\bibinfo{year}{1998}).

\bibitem[{\citenamefont{Ma{\'c}kowiak and Tarasewicz}(2000)}]{Mackowiak2000A}
\bibinfo{author}{\bibfnamefont{J.}~\bibnamefont{Ma{\'c}kowiak}}
  \bibnamefont{and}
  \bibinfo{author}{\bibfnamefont{P.}~\bibnamefont{Tarasewicz}},
  \bibinfo{journal}{Physica C} \textbf{\bibinfo{volume}{331}},
  \bibinfo{pages}{25} (\bibinfo{year}{2000}).

\bibitem[{\citenamefont{Ma{\'c}kowiak and Baran}(2011)}]{Mackowiak2011A}
\bibinfo{author}{\bibfnamefont{J.}~\bibnamefont{Ma{\'c}kowiak}}
  \bibnamefont{and} \bibinfo{author}{\bibfnamefont{D.}~\bibnamefont{Baran}},
  \bibinfo{journal}{International Journal of Modern Physics B}
  \textbf{\bibinfo{volume}{25}}, \bibinfo{pages}{1701} (\bibinfo{year}{2011}).

\bibitem[{\citenamefont{Szcz{\c{e}}{\'s}niak}(2012{\natexlab{b}})}]{Szczesniak2012H}
\bibinfo{author}{\bibfnamefont{R.}~\bibnamefont{Szcz{\c{e}}{\'s}niak}},
  \bibinfo{journal}{xxx.lanl.gov} \textbf{\bibinfo{volume}{preprint:
  arXiv:1110.3404}}, \bibinfo{pages}{1} (\bibinfo{year}{2012}{\natexlab{b}}).

\bibitem[{\citenamefont{Szcz{\c{e}}{\'s}niak and
  Durajski}(2014{\natexlab{a}})}]{Szczesniak2014C}
\bibinfo{author}{\bibfnamefont{R.}~\bibnamefont{Szcz{\c{e}}{\'s}niak}}
  \bibnamefont{and} \bibinfo{author}{\bibfnamefont{A.~P.}
  \bibnamefont{Durajski}}, \bibinfo{journal}{Superconductor Science and
  Technology} \textbf{\bibinfo{volume}{27}}, \bibinfo{pages}{125004}
  (\bibinfo{year}{2014}{\natexlab{a}}).

\bibitem[{\citenamefont{Szcz{\c{e}}{\'s}niak and
  Durajski}(2014{\natexlab{b}})}]{Szczesniak2014D}
\bibinfo{author}{\bibfnamefont{R.}~\bibnamefont{Szcz{\c{e}}{\'s}niak}}
  \bibnamefont{and} \bibinfo{author}{\bibfnamefont{A.~P.}
  \bibnamefont{Durajski}}, \bibinfo{journal}{Journal of Superconductivity and
  Novel Magnetism} \textbf{\bibinfo{volume}{27}}, \bibinfo{pages}{1363}
  (\bibinfo{year}{2014}{\natexlab{b}}).

\bibitem[{\citenamefont{Szcz{\c{e}}{\'s}niak and
  Durajski}(2014{\natexlab{c}})}]{Szczesniak2014E}
\bibinfo{author}{\bibfnamefont{R.}~\bibnamefont{Szcz{\c{e}}{\'s}niak}}
  \bibnamefont{and} \bibinfo{author}{\bibfnamefont{A.~P.}
  \bibnamefont{Durajski}}, \bibinfo{journal}{Acta Physica Polonica A}
  \textbf{\bibinfo{volume}{126}}, \bibinfo{pages}{A92}
  (\bibinfo{year}{2014}{\natexlab{c}}).

\bibitem[{\citenamefont{Szcz{\c{e}}{\'s}niak and
  Durajski}(2014{\natexlab{d}})}]{Szczesniak2014H}
\bibinfo{author}{\bibfnamefont{R.}~\bibnamefont{Szcz{\c{e}}{\'s}niak}}
  \bibnamefont{and} \bibinfo{author}{\bibfnamefont{A.~P.}
  \bibnamefont{Durajski}}, \bibinfo{journal}{Journal of Superconductivity and
  Novel Magnetism} \textbf{\bibinfo{volume}{28}}, \bibinfo{pages}{19}
  (\bibinfo{year}{2014}{\natexlab{d}}).

\bibitem[{\citenamefont{Szcz{\c{e}}{\'s}niak
  et~al.}(2015)\citenamefont{Szcz{\c{e}}{\'s}niak, Jarosik, and
  Duda}}]{Szczesniak2015C}
\bibinfo{author}{\bibfnamefont{R.}~\bibnamefont{Szcz{\c{e}}{\'s}niak}},
  \bibinfo{author}{\bibfnamefont{M.~W.} \bibnamefont{Jarosik}},
  \bibnamefont{and} \bibinfo{author}{\bibfnamefont{A.~M.} \bibnamefont{Duda}},
  \bibinfo{journal}{Advances in Condensed Matter Physics}
  \textbf{\bibinfo{volume}{2015}}, \bibinfo{pages}{1} (\bibinfo{year}{2015}).

\bibitem[{\citenamefont{Szcz{\c{e}}{\'s}niak}(2005)}]{Szczesniak2005A}
\bibinfo{author}{\bibfnamefont{R.}~\bibnamefont{Szcz{\c{e}}{\'s}niak}},
  \bibinfo{journal}{Physics Letters A} \textbf{\bibinfo{volume}{336}},
  \bibinfo{pages}{402} (\bibinfo{year}{2005}).

\bibitem[{\citenamefont{Szcz{\c{e}}{\'s}niak and
  Grabi{\'n}ski}(2002)}]{Szczesniak2002A}
\bibinfo{author}{\bibfnamefont{R.}~\bibnamefont{Szcz{\c{e}}{\'s}niak}}
  \bibnamefont{and}
  \bibinfo{author}{\bibfnamefont{S.}~\bibnamefont{Grabi{\'n}ski}},
  \bibinfo{journal}{Acta Physica Polonica A} \textbf{\bibinfo{volume}{102}},
  \bibinfo{pages}{401} (\bibinfo{year}{2002}).

\bibitem[{\citenamefont{Gofron et~al.}(1994)\citenamefont{Gofron, Campuzano,
  Abrikosov, Lindroos, Bansil, Ding, Koelling, and Dabrowski}}]{Gofron1994A}
\bibinfo{author}{\bibfnamefont{K.}~\bibnamefont{Gofron}},
  \bibinfo{author}{\bibfnamefont{J.~C.} \bibnamefont{Campuzano}},
  \bibinfo{author}{\bibfnamefont{A.~A.} \bibnamefont{Abrikosov}},
  \bibinfo{author}{\bibfnamefont{M.}~\bibnamefont{Lindroos}},
  \bibinfo{author}{\bibfnamefont{A.}~\bibnamefont{Bansil}},
  \bibinfo{author}{\bibfnamefont{H.}~\bibnamefont{Ding}},
  \bibinfo{author}{\bibfnamefont{D.}~\bibnamefont{Koelling}}, \bibnamefont{and}
  \bibinfo{author}{\bibfnamefont{B.}~\bibnamefont{Dabrowski}},
  \bibinfo{journal}{Physical Review Letters} \textbf{\bibinfo{volume}{73}},
  \bibinfo{pages}{3302} (\bibinfo{year}{1994}).

\bibitem[{\citenamefont{Szcz{\c{e}}{\'s}niak}(2006{\natexlab{a}})}]{Szczesniak2006B}
\bibinfo{author}{\bibfnamefont{R.}~\bibnamefont{Szcz{\c{e}}{\'s}niak}},
  \bibinfo{journal}{Acta Physica Polonica A} \textbf{\bibinfo{volume}{109}},
  \bibinfo{pages}{179} (\bibinfo{year}{2006}{\natexlab{a}}).

\bibitem[{\citenamefont{Szcz{\c{e}}{\'s}niak}(2006{\natexlab{b}})}]{Szczesniak2006A}
\bibinfo{author}{\bibfnamefont{R.}~\bibnamefont{Szcz{\c{e}}{\'s}niak}},
  \bibinfo{journal}{Solid State Communications} \textbf{\bibinfo{volume}{138}},
  \bibinfo{pages}{347} (\bibinfo{year}{2006}{\natexlab{b}}).

\bibitem[{\citenamefont{Liang et~al.}(2006)\citenamefont{Liang, Bonn, and
  Hardy}}]{Liang2006A}
\bibinfo{author}{\bibfnamefont{R.}~\bibnamefont{Liang}},
  \bibinfo{author}{\bibfnamefont{D.~A.} \bibnamefont{Bonn}}, \bibnamefont{and}
  \bibinfo{author}{\bibfnamefont{W.~N.} \bibnamefont{Hardy}},
  \bibinfo{journal}{Physical Review B} \textbf{\bibinfo{volume}{73}},
  \bibinfo{pages}{180505(R)} (\bibinfo{year}{2006}).

\bibitem[{\citenamefont{Padilla et~al.}(2005)\citenamefont{Padilla, Lee, Dumm,
  Blumberg, Ono, Segawa, Komiya, Ando, and Basov}}]{Padilla2005A}
\bibinfo{author}{\bibfnamefont{W.~J.} \bibnamefont{Padilla}},
  \bibinfo{author}{\bibfnamefont{Y.~S.} \bibnamefont{Lee}},
  \bibinfo{author}{\bibfnamefont{M.}~\bibnamefont{Dumm}},
  \bibinfo{author}{\bibfnamefont{G.}~\bibnamefont{Blumberg}},
  \bibinfo{author}{\bibfnamefont{S.}~\bibnamefont{Ono}},
  \bibinfo{author}{\bibfnamefont{K.}~\bibnamefont{Segawa}},
  \bibinfo{author}{\bibfnamefont{S.}~\bibnamefont{Komiya}},
  \bibinfo{author}{\bibfnamefont{Y.}~\bibnamefont{Ando}}, \bibnamefont{and}
  \bibinfo{author}{\bibfnamefont{D.~N.} \bibnamefont{Basov}},
  \bibinfo{journal}{Physical Review B} \textbf{\bibinfo{volume}{72}},
  \bibinfo{pages}{060511(R)} (\bibinfo{year}{2005}).

\bibitem[{\citenamefont{Deemyad and Schilling}(2003)}]{Deemyad2003A}
\bibinfo{author}{\bibfnamefont{S.}~\bibnamefont{Deemyad}} \bibnamefont{and}
  \bibinfo{author}{\bibfnamefont{J.~S.} \bibnamefont{Schilling}},
  \bibinfo{journal}{Physical Review Letters} \textbf{\bibinfo{volume}{91}},
  \bibinfo{pages}{167001} (\bibinfo{year}{2003}).

\bibitem[{\citenamefont{Yabuuchi et~al.}(2006)\citenamefont{Yabuuchi, Matsuoka,
  Nakamoto, and Shimizu}}]{Yabuuchi2006A}
\bibinfo{author}{\bibfnamefont{T.}~\bibnamefont{Yabuuchi}},
  \bibinfo{author}{\bibfnamefont{T.}~\bibnamefont{Matsuoka}},
  \bibinfo{author}{\bibfnamefont{Y.}~\bibnamefont{Nakamoto}}, \bibnamefont{and}
  \bibinfo{author}{\bibfnamefont{K.}~\bibnamefont{Shimizu}},
  \bibinfo{journal}{Journal of the Physical Society of Japan}
  \textbf{\bibinfo{volume}{75}}, \bibinfo{pages}{083703}
  (\bibinfo{year}{2006}).

\bibitem[{\citenamefont{Sakata et~al.}(2011)\citenamefont{Sakata, Nakamoto,
  Shimizu, Matsuoka, and Ohishi}}]{Sakata2011A}
\bibinfo{author}{\bibfnamefont{M.}~\bibnamefont{Sakata}},
  \bibinfo{author}{\bibfnamefont{Y.}~\bibnamefont{Nakamoto}},
  \bibinfo{author}{\bibfnamefont{K.}~\bibnamefont{Shimizu}},
  \bibinfo{author}{\bibfnamefont{T.}~\bibnamefont{Matsuoka}}, \bibnamefont{and}
  \bibinfo{author}{\bibfnamefont{Y.}~\bibnamefont{Ohishi}},
  \bibinfo{journal}{Physical Review B} \textbf{\bibinfo{volume}{83}},
  \bibinfo{pages}{220512(R)} (\bibinfo{year}{2011}).

\bibitem[{\citenamefont{Strobel et~al.}(2010)\citenamefont{Strobel, Chen,
  Somayazulu, and Hemley}}]{Strobel2010A}
\bibinfo{author}{\bibfnamefont{T.~A.} \bibnamefont{Strobel}},
  \bibinfo{author}{\bibfnamefont{X.~J.} \bibnamefont{Chen}},
  \bibinfo{author}{\bibfnamefont{M.}~\bibnamefont{Somayazulu}},
  \bibnamefont{and} \bibinfo{author}{\bibfnamefont{R.~J.}
  \bibnamefont{Hemley}}, \bibinfo{journal}{The Journal of Chemical Physics}
  \textbf{\bibinfo{volume}{133}}, \bibinfo{pages}{164512}
  (\bibinfo{year}{2010}).

\bibitem[{\citenamefont{Eremets et~al.}(2008)\citenamefont{Eremets, Trojan,
  Medvedev, Tse, and Yao}}]{Eremets2008A}
\bibinfo{author}{\bibfnamefont{M.~I.} \bibnamefont{Eremets}},
  \bibinfo{author}{\bibfnamefont{I.~A.} \bibnamefont{Trojan}},
  \bibinfo{author}{\bibfnamefont{S.~A.} \bibnamefont{Medvedev}},
  \bibinfo{author}{\bibfnamefont{J.~S.} \bibnamefont{Tse}}, \bibnamefont{and}
  \bibinfo{author}{\bibfnamefont{Y.}~\bibnamefont{Yao}},
  \bibinfo{journal}{Science} \textbf{\bibinfo{volume}{319}},
  \bibinfo{pages}{1506} (\bibinfo{year}{2008}).

\bibitem[{\citenamefont{Beach et~al.}(2000)\citenamefont{Beach, Gooding, and
  Marsiglio}}]{Beach2000A}
\bibinfo{author}{\bibfnamefont{K.~S.~D.} \bibnamefont{Beach}},
  \bibinfo{author}{\bibfnamefont{R.~J.} \bibnamefont{Gooding}},
  \bibnamefont{and}
  \bibinfo{author}{\bibfnamefont{F.}~\bibnamefont{Marsiglio}},
  \bibinfo{journal}{Physical Review B} \textbf{\bibinfo{volume}{61}},
  \bibinfo{pages}{5147} (\bibinfo{year}{2000}).

\bibitem[{\citenamefont{Varelogiannis}(1997)}]{Varelogiannis1997A}
\bibinfo{author}{\bibfnamefont{G.}~\bibnamefont{Varelogiannis}},
  \bibinfo{journal}{Zeitschrift f{\" u}r Physik B Condensed Matter}
  \textbf{\bibinfo{volume}{104}}, \bibinfo{pages}{411} (\bibinfo{year}{1997}).

\end{thebibliography}
\end{document}